\documentclass[a4paper,11pt]{article}
\pdfoutput=1 

\usepackage{jcappub} 

\usepackage[T1]{fontenc} 
\usepackage{subcaption}

\title{\boldmath Rapid Simulations of Halo and Subhalo Clustering}

\author[a,1]{Pascale Berner,\note{Corresponding author.}}
\author[a]{Alexandre Refregier,}
\author[a]{Raphael Sgier,}
\author[a, b]{Tomasz Kacprzak,}
\author[a, c]{Luca Tortorelli}
\author[d, e, f, g]{and Pierluigi Monaco}

\affiliation[a]{Institute for Particle Physics and Astrophysics, ETH Z\"urich, 8092 Zurich, Switzerland}
\affiliation[b]{Swiss Data Science Center, Paul Scherrer Institute, 5232 Villigen, Switzerland}
\affiliation[c]{University Observatory, Faculty of Physics, Ludwig-Maximilians-Universit\"at M\"unchen, \\ 81679 Munich, Germany}
\affiliation[d]{Dipartimento di Fisica, Universita di Trieste, 34143 Trieste, Italy}
\affiliation[e]{Istituto Nazionale di Astrofisica, Osservatorio Astronomico di Trieste, 34143 Trieste, Italy}
\affiliation[f]{Institute for Fundamental Physics of the Universe, 34014 Trieste, Italy}
\affiliation[g]{Istituto Nazionale di Fisica Nucleare, Sezione di Trieste, 34127 Trieste, Italy}

\emailAdd{pascale.berner@phys.ethz.ch}

\abstract{
The analysis of cosmological galaxy surveys requires realistic simulations for their interpretation. Forward modelling is a powerful method to simulate galaxy clustering without the need for an underlying complex model. This approach requires fast cosmological simulations with a high resolution and large volume, to resolve small dark matter halos associated to single galaxies. In this work, we present fast halo and subhalo clustering simulations based on the Lagrangian perturbation theory code \texttt{PINOCCHIO}, which generates halos and merger trees. The subhalo progenitors are extracted from the merger history and the survival of subhalos is modelled. We introduce a new fitting function for the subhalo merger time, which includes a redshift dependence of the fitting parameters. The spatial distribution of subhalos within their hosts is modelled using a number density profile. We compare our simulations with the halo finder \texttt{ROCKSTAR} applied to the full N-body code \texttt{GADGET-2}. The subhalo velocity function and the correlation function of halos and subhalos are in good agreement. We investigate the effect of the chosen number density profile on the resulting subhalo clustering. Our simulation is approximate yet realistic and significantly faster compared to a full N-body simulation combined with a halo finder. The fast halo and subhalo clustering simulations offer good prospects for galaxy forward models using subhalo abundance matching.
}

\begin{document}
\maketitle
\flushbottom

\section{Introduction}
\label{sec:intro}
Current and future galaxy surveys, both imaging and spectroscopic, cover larger and larger volumes of the Universe. Given their increasing sensitivity and depth, their design and evaluation require large simulations. Since full hydrodynamical simulations are complex and computationally expensive, simulating galaxies and their spatial distribution directly is difficult. Pure dark matter simulations are better understood and more efficient, since less physical processes have to be included. Gravity being the only force affecting dark matter allows the usage of simulation particles instead of actual physical particles. Since baryonic matter falls into over-densities of the dark matter field, galaxies tend to reside in regions of high dark matter density.\\
A powerful way to distribute galaxies in a dark matter simulation is by associating them with dark matter halos. Halos are collapsed structures of dark matter. Most dark matter and therefore most simulation particles can be assigned to dark matter halos. They trace the dark matter density field and make up the densest regions in space. Through hierarchical structure formation halos of different masses form, whereas more massive halos are more strongly clustered than lighter ones, meaning their clustering signal is stronger. Baryons fall into the gravitational wells of dark matter; therefore, galaxies form in regions of high dark matter density. Consequently, the positions of galaxies can be associated to those of dark matter halos.\\
The most common methods to populate halos with galaxies are with Halo Occupation distribution HOD (e.g. \cite{Jing_1998, Seljak_2000, Peacock_2000, Scoccimarro_2001, hod_2002, cooray_2002, hod_2005}), (Sub-)Halo Abundance Matching (S)HAM (e.g. \cite{Kravtsov_2004, Vale_2004, Conroy_2006, behroozi_2010, sham_2012, sham_2013, Guo_2016}) or semi-analytic models (e.g. \cite{2000_cole, 2008_somerville, 2011_guo}, \texttt{GAEA} \cite{2004gaea, 2014gaea, 2016gaea, 2017gaea}). Semi-analytic models are powerful but require modelling galaxy clustering and formation, which makes the models relatively complicated. HOD calls for the introduction of two functions N$_{\mathrm{cen}}$ and N$_{\mathrm{sat}}$, the average number of central and satellite galaxies in a halo as a function of halo mass and redshift. Halo Abundance Matching on the other hand assigns only one galaxy to each halo by matching the number of objects for example by halo mass and galaxy luminosity. As a result, SHAM does not need these two functions and it gives the advantage of having positions and velocities for the galaxies.\\
Since each halo can only host one galaxy in a HAM model, the resolution of the simulation must be good enough for halos hosting single galaxies to be resolved. Depending on the lower limit in luminosity of the galaxies that should be included in the simulation, rather small halos are needed. Furthermore, satellite galaxies as well as galaxies in a cluster are hosted by subhalos \cite{2009yang}. Therefore, one uses Subhalo Abundance Matching and needs a catalogue with both halos and subhalos.\\
While the accuracy of a simulation can be important, it usually coincides with high computational cost and often complicated models. For many applications, fast simulations are essential, allowing also for approximate methods. Examples for such applications are Approximate Bayesian Computation (ABC, e.g. \cite{2013weyant, Akeret_2015}) or other approaches for Likelihood-Free Inference. A large number of simulations is typically needed, potentially requiring furthermore that parameters and other settings are selected on the fly. This can make it difficult to run a fixed grid of simulations in advance.\\
Different dark matter N-body simulation codes are publicly available, such as \texttt{GADGET-2} \cite{gadget2001, gadget2005}, \texttt{GADGET-4} \cite{gadget4_2021}, \texttt{PKDGRAV3} \cite{potter2016pkdgrav3} and \texttt{ABACUS} \cite{abacus_2019}. The most common method to identify dark matter halos is to run a halo finder on a dark matter particle simulation, e.g. \texttt{ROCKSTAR} \cite{behroozi_2012, Behroozi_2019} or \texttt{AHF} \cite{2009_AHF}. The dependence of halo properties on the halo finder used has been studied in e.g. \cite{2011_knebe}. An extensive study presented in \cite{2022_massfunctioncalibration} discusses the calibration of the halo mass function and is based on multiple suits of simulations using different codes. Zoom-in simulations have been used to study the substructure within dark matter halos, including the subhalos and their distribution. There are studies describing the evolution and survival of subhalos within their hosts over time (e.g. \cite{1987binney, 1993lacey, 1998tormen, boylon-kolchlin2008, Jiang_2008, 2010_jiang, 2009fakhouri, 2009stewart, 2010hester, 2017poole}) as well as the spatial distribution of substructures (e.g. \cite{Zentner_2005, Diemand_2004, 2004gao, 2006zentner, 2016han}).\\
Beside the accurate but computationally expensive N-body simulations there exist also some approximate methods to generate dark matter halo catalogues (summarised in e.g. \cite{2016appsim}). To name only a few, there are for example \texttt{PINOCCHIO} \cite{pinocchio2002, pinocchio2013, pinocchio2017}, \texttt{HALOGEN} \cite{halogen2015} and \texttt{L-PICOLA} \cite{howlett2015lpicola}. \texttt{HALOGEN} requires a particle snapshot for the dark matter distribution and then paints on dark matter halos based on a calibration with an accurate halo catalogue. It constructs the matter density field with second order Lagrangian Perturbation Theory (2LPT). \texttt{L-PICOLA} is based on \texttt{COLA} \cite{cola2013}, which is a Particle Mesh code that computes the displacements from the 2LPT trajectory, such that the convergence on large scales is obtained with few timesteps. It can be combined with a halo finder. \texttt{PINOCCHIO} starts from the density field sampled on a grid, as the N-body simulations, and gives halo catalogues and a merger history directly. It is described in more detail in section \ref{sec:method_distribution}. Further approximate methods to generate mock catalogs include \texttt{PeakPATCH} \cite{1996apeakpatch, 1996bpeakpatch, 1996cpeakpatch, 2019peakpatch}, \texttt{FastPM} \cite{2016fastpm} and \texttt{BAM} \cite{2019bam}. \texttt{PTHalos} \cite{pthalos2001}, \texttt{PATCHY} \cite{patchy2013}, \texttt{QPM} \cite{qpm2014} and \texttt{EZmocks} \cite{2015ezmocks} were used for the galaxy surveys BOSS \cite{sdssIII} and eBOSS \cite{sdssIV}.\\
In this paper, we develop a fast simulation technique to simulate dark matter halos and subhalos, with realistic properties and clustering, that can be used, for example, for subhalo abundance matching. Galaxy models can then be constructed using a forward modelling approach, requiring a simple model that can be run fast and many times (see e.g. \cite{Fagioli_2020} for a discussion on forward modelling). Starting from \texttt{PINOCCHIO} for the halos and the merger history, we derived the surviving subhalos for a snapshot or lightcone simulation with halos and subhalos, using a new scheme for the subhalo survival time. We thus generated a fast simulation that is not based on a complicated theoretical model. This work describes the model for the halo and subhalo simulation. In a future work, we will implement the SHAM method to create clustered galaxy simulations (P. Berner et al., in preparation). Our simulations can also be used for modelling cosmological neutral hydrogen on small scales efficiently.\\ 
The outline of this paper is as follows. In section \ref{sec:method} we describe our method for extracting subhalos from the merger history of \texttt{PINOCCHIO} as well as our model for the survival of subhalos. This section also includes a description of how we compared our results to a full N-body simulation with a halo finder. We present our best fit for the subhalo survival time for \texttt{PINOCCHIO} subhalos in section \ref{sec:results} followed by the resulting subhalo velocity function and the clustering properties of our final simulation. Finally in section \ref{sec:conc} we summarise our findings and outline future steps.

\section{Method}
\label{sec:method}
In this section, we describe how we derive the surviving subhalos from a merger history and how we determine their position within their host halo. We also compared the properties of the resulting halo-subhalo catalogue to a catalogue from a halo finder run on a full N-body simulation.

\subsection{Subhalos from a merger tree}
\label{sec:method_merger_tree}
Structure formation with only dark matter is well modelled with N-body simulations. Full N-body simulations are computationally expensive to run and are therefore not ideal for a fast forward modelling approach. Additionally, we need a relatively high resolution to resolve objects corresponding to single galaxies, ideally corresponding to halo masses down to about $10^9$ M$_\odot$/h. Furthermore, large volumes must be simulated to realistically model wide-field surveys, which leads to even higher computational costs for simulations of a given resolution. Approximate dark matter simulations take less computational time to run at a trade-off of precision, typically on small scales given a certain resolution (e.g. \texttt{L-PICOLA} \cite{howlett2015lpicola}). Here, we use version 4.1.3 of \texttt{PINOCCHIO} \cite{pinocchio2002, pinocchio2013, pinocchio2017} as the main dark matter simulation tool.

\begin{figure}[b]
\centering 
\includegraphics[width=.80\textwidth,angle=0]{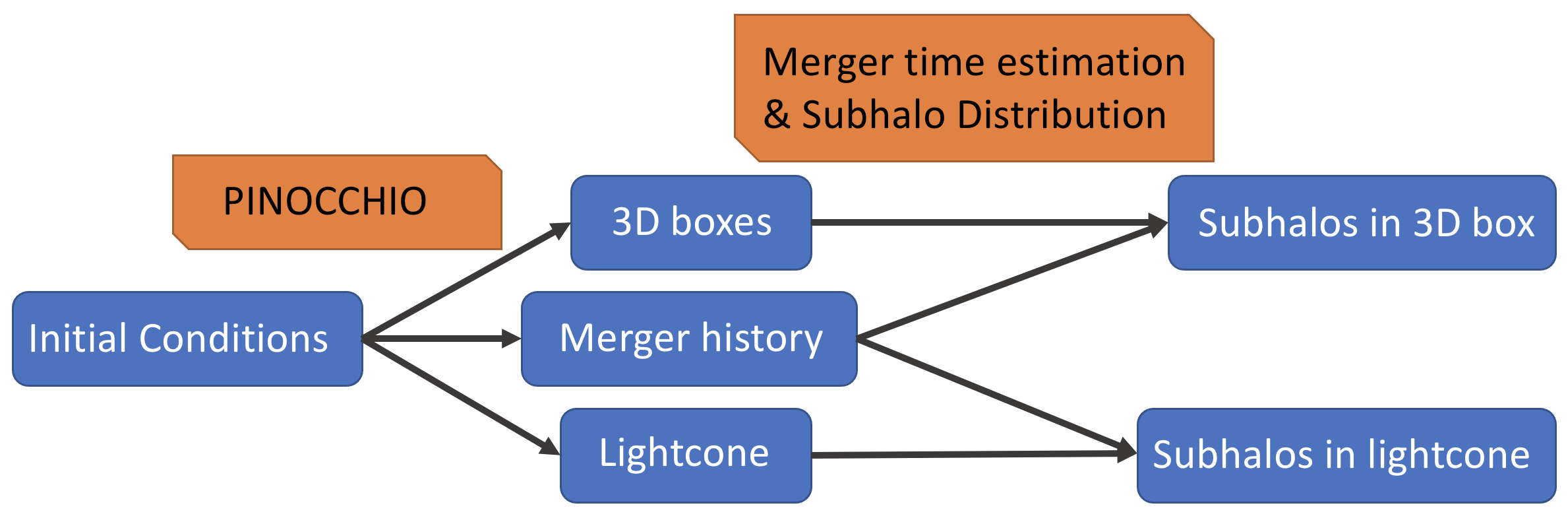}
\caption{Schematic of the method to generate  halo-subhalo catalogues from the \texttt{PINOCCHIO} output, for both snapshots and lightcones.}
\label{fig:pin_subs}
\end{figure} 

\noindent \texttt{PINOCCHIO} combines Lagrangian Perturbation Theory (LPT) and the Extended Press \& Schechter Formalism (EPS) \cite{1991_EPS}, making it an approximation to a full N-body simulation combined with a halo finder run at each time step. The starting point of the simulation are initial conditions on a cubic grid. The loop of \texttt{N-GenIC} \cite{ngenic2005, ngenic2012} that creates the linear density field is implemented inside \texttt{PINOCCHIO}. At first \texttt{PINOCCHIO} computes the collapse time for each particle using ellipsoidal collapse, in the way of orbit crossing. The fragmentation process then decides, at each time step, whether a collapsed particle belongs to a halo or the filament network. For this, the six neighbouring particles and their states are considered. \texttt{PINOCCHIO} therefore combines particles within collapsed structures into groups, resulting in a strong speed-up at the cost of information on halo substructure. After a merger, information on the mass, position, or velocity of the smaller group prior to merging is no longer stored. The speed-up results from the decreasing number of objects that interact gravitationally and must be traced as the simulation progresses. \texttt{PINOCCHIO} is predictive on the mass function and is calibrated to obtain the mass function of Friends-of-Friends halos. \texttt{PINOCCHIO} directly outputs halo catalogues at desired redshifts, an on-the-fly generated halo lightcone of a chosen width and depth and the merger history of dark matter particle groups for the halos in the final snapshot. The merger history contains the information on every merger that happened for particle groups above a specified resolution and is organised based on the final snapshot. The outputs of \texttt{PINOCCHIO} are schematically shown in the middle of figure \ref{fig:pin_subs}.\\
The performance of \texttt{PINOCCHIO} is well documented \cite{pinocchio2002, pinocchio2013, pinocchio2017}. The speed-up compared to full N-body simulations was tested, both for different resolutions and box sizes. Furthermore, the accuracy of the halo mass function and of the clustering of halos was assessed.\\
Large dark matter halos host baryonic objects like large galaxy clusters, consisting of many galaxies within the same dark matter halo. To use abundance matching it is therefore not enough to have a catalogue of halos, but the subhalos are needed additionally. While \texttt{PINOCCHIO} is made for fast simulations down to good resolutions without having to reduce the volume, it does not include subhalos in the final catalogues. Part of the speed-up in \texttt{PINOCCHIO} is that it does not compute or save substructure within a halo. We can instead estimate the surviving subhalos using the merger history of particle groups.\\
The information available from the merger history of PINOCCHIO includes the redshift at the start of merger, the mass of the substructure and of the host at this time, IDs uniquely identifying both the substructure and its host, as well as temporal information on the formation of the future subhalo. The start of merger is defined here as the time when \texttt{PINOCCHIO} starts treating the host and the indicated substructure as one halo. No information on the orbital velocity of the substructure at the start of merger is known from the merger history.\\
Since the merger timescale does not give the exact mass loss as a function of time, the only mass known for the surviving subhalos is the accretion mass. The detected mass at the observed redshift would be lower. For the catalogue that we create here, we ignore the mass loss of the surviving subhalos. This does however mean that we overestimate the mass of the subhalos. Some subhalos are assigned to the wrong mass bins when we look at the halo and subhalo clustering, but we only do that for checking our model. The accretion mass of the surviving subhalos is well suited as a matching parameter for subhalo abundance matching (see e.g. \cite{Wechsler_2018}), so we do not estimate the mass loss after the accretion here. The mass of substructures is not well defined, and galaxy properties depend on the mass at accretion time, not on the stripped mass. It is therefore convenient to use the mass at accretion time and there is no real need to predict the final accreted mass. Our simulated value of the subhalo mass hence does not evolve until the subhalo stops existing.\\
For the host halos we use the simulated mass at the observed redshift. This mass includes the mass contained in its subhalos, so the subhalo masses are double counted. Therefore, the total mass in all halos and subhalos in our resulting catalogue may be more than the mass expected in the corresponding volume. Some mass can also be unaccounted for if particles are in halos below the specified mass limit. This diffuse mass depends on the resolution.\\
The merger history from the output in \texttt{PINOCCHIO} is written for all halos at the last calculated redshift. At higher redshifts, some halos are hosts that merge into larger halos later on in the simulation. They become subhalos or indistinguishable substructure until the final redshift. These host halos are therefore not directly covered as halos in the merger history. To calculate the subhalos of hosts at higher redshifts (in snapshots at higher redshifts or in the lightcone), we trace back each merger in the merger history to find all the subgroups. For each particle group, we get a list of groups that merge into it at any redshift during the simulation. This is a lot more efficient than outputting the merger history for multiple snapshots. We can then estimate the merger timescale for all the subgroups inside a halo at any redshift. The merger time estimation and the way we distribute the surviving subhalos is schematically shown on the right side of figure \ref{fig:pin_subs}.

\subsection{Model for the subhalo survival time}
\label{sec:method_survival_time}
After merging with a larger halo, a halo becomes a subhalo and interacts with the remaining structure inside the host halo. Processes affecting the subhalos include dynamical friction, tidal stripping and tidal heating (e.g. \cite{Bosch2005, Gan_2010}). A subhalo survives in our model if the time difference between accretion and observation, calculated from the accretion redshift in the merger history and the redshift of the host halo in the lightcone or snapshot, is smaller than the estimated merger timescale. We define the merger time as the duration it takes for a given substructure to get stripped enough within the host to not be treated as an individual collapsed structure anymore. The merger time is therefore the time interval between the start of merger and the time when the subhalo is not detected anymore.\\
Other works (e.g. \cite{boylon-kolchlin2008, Jiang_2008, 2012mccavana, 2013villalobos}) have presented different fitting formulas for the merger timescale, also called subhalo survival time. Generally, the models describe functions of the following form:
\begin{equation}
    \tau_{\mathrm{merge}} = f(\tau_{\mathrm{dyn}}, M_{\mathrm{host}}, m_{\mathrm{sub}}, \eta, V_c, r_c, r_{\mathrm{vir}}, z)
\end{equation}
with potentially other dependencies on orbital parameters. The main dependence is on the host and subhalo masses $M_{\mathrm{host}}$ and $m_{\mathrm{sub}}$ at merger. They typically appear only as a ratio, meaning as $(M_{\mathrm{host}}/m_{\mathrm{sub}})$. We get these masses at the time of accretion directly from the merger history of \texttt{PINOCCHIO}.\\
The timescale $\tau_{\mathrm{dyn}}$ is the dynamical time at the virial radius $r_{\mathrm{vir}}$ of the host halo and can be derived directly from the Hubble constant H, which in term depends on the Cosmological parameters:
\begin{equation}
   \tau_{\mathrm{dyn}} \approx 0.1 H^{-1} 
\end{equation}
The remaining parameters describe the orbit of the subhalo within the host, where $\eta$ is the orbital circulariy, $V_c$ the circular velocity, $r_c$ the radius for a circular orbit and $r_{\mathrm{vir}}$ the virial radius of the host halo for scaling.\\
In \cite{boylon-kolchlin2008}, a model of the following form is presented
\begin{equation}
    \tau_{\mathrm{merge}} = A ~ \tau_{\mathrm{dyn}} ~  \frac{(M_{\mathrm{host}}/m_{\mathrm{sub}})^b}{\ln(1+M_{\mathrm{host}}/m_{\mathrm{sub}})} \exp \big(c \eta \big) \bigg(\frac{r_c}{r_{\mathrm{vir}}}\bigg)^d ~,
\label{eqn:kolchlin_fit}
\end{equation}
where the best fit parameters are $A=0.216$, $b=1.3$, $c=1.9$ and $d=1.0$. The authors describe a reduction of the merging timescale by 10\% (meaning an additional factor of $0.9$ multiplied to $A$) if baryonic bulges at the galactic centre are included.\\
In \cite{Jiang_2008} a similar mass dependence is shown, but with a different dependence on the orbital parameters:
\begin{equation}
    \tau_{\mathrm{merge}} = A ~  \frac{(M_{\mathrm{host}}/m_{\mathrm{sub}})}{\ln(1+M_{\mathrm{host}}/m_{\mathrm{sub}})} \frac{r_{\mathrm{vir}}}{V_c} \big(\eta^{\alpha} + \alpha \big) ~,
\end{equation}
with $\alpha = 0.60$. For better readability, we have relabelled the parameters from the description in \cite{Jiang_2008} to match our definitions. The mass dependence has the same functional form as the one in the previous fitting function by \cite{boylon-kolchlin2008}, only with $b=1$.\\
Since \texttt{PINOCCHIO} does not calculate any information on the trajectory of the subhalo after their merger, we do not have any orbital parameters for the subhalos within their hosts. While we can get a good estimate for the virial radius of the host halo knowing its mass, assuming an NFW mass density profile (Nawarro, Frenk and White \cite{nfw_1997}), we cannot estimate the orbital parameters of the subhalos easily. In principle it may be possible to use the velocities of halos to infer the orbital parameters. This could be investigated in future work.\\
The ratio $(r_c/r_{\mathrm{vir}})$ corresponds to the orbital energy. It can be drawn from a uniform distribution in the interval $[0.1, 1]$. We finally chose to omit this term instead, effectively setting $d$ to 0, since a uniformly drawn orbital energy statistically only affects the normalization of the merger time.\\
The orbital circularity $\eta$ is drawn from the following distribution shown by \cite{Zentner_2005}:
\begin{equation}
    P(\eta) \propto \eta^{1.2} (1-\eta)^{1.2}
\end{equation}
from the interval $[0.2, 1]$, as explained in \cite{Birrer_2014}. We do not need the normalization of this distribution here, since we only sample from it.\\
As discussed in more detail in section \ref{sec:method_nbody}, subhalo masses are affected by dynamical processes of tidal stripping and tidal shocks that are difficult to represent at the relatively low resolutions used in this paper, so their values do evolve in time and are affected by numerical effects. A more stable quantity is the maximum of the subhalo "rotation curve" $V=\sqrt{GM/r}$, that is commonly called $V_{\rm max}$; this maximum lies in the internal regions of the halo that are less affected by tidal processes, so $V_{\rm max}$ is more stable along the subhalo history, and thus more representative of the halo mass at accretion time. We thus relied for our analysis on $V_{\rm max}$, instead of subhalo mass, to match \texttt{PINOCCHIO} and N-body halo statistics. With the above mentioned fitting functions by \cite{boylon-kolchlin2008} and \cite{Jiang_2008}, too many subhalos are surviving, especially in halos of high masses. This results in an overabundance of subhalos compared to halos, which would be a problem for an application like subhalo abundance matching. Furthermore, too many surviving subhalos in high mass halos results in an overestimation of the clustering of subhalos.\\
Since the main dependence on halo and subhalo mass should be on the ratio between the two, we kept the functional form from equation (\ref{eqn:kolchlin_fit}). We tested the effect of the different fitting parameters on the resulting subhalo velocity function for snapshots at different redshifts. We investigated different values of $A$ and $b$ on a grid and found best fit values for each tested redshift. To minimize the dependence of the fitting parameters on the chosen cosmological parameters, we fitted the redshift dependence of the parameters as a function of the linear growth rate factor $D(z)$. Since $D(z)$ is the driver of structure formation, such a parametrization guarantees a better stability of parameter values with cosmology. It is often written $D(a)$ as a function of scale factor. Here we assume that the linear growth rate $D(z)$ is normalized to unity at $z=0$, as the scale factor.\\
We aimed for a dependence on $D(z)$ that is as simple as possible. This gives us the following fitting function for the subhalo survival time, for which we show our best fit parameters in section \ref{sec:res_fit}:
\begin{equation}
    \tau_{\mathrm{merge}} = A(D) ~ \tau_{\mathrm{dyn}} ~  \frac{(M_{\mathrm{host}}/m_{\mathrm{sub}})^{b(D)}}{\ln(1+M_{\mathrm{host}}/m_{\mathrm{sub}})} \exp \big(c \eta \big)
\label{eqn:merger_time_form}
\end{equation}

\subsection{Subhalo distribution within hosts}
\label{sec:method_distribution}
\begin{figure}[t]
\centering 
\includegraphics[width=.68\textwidth,angle=0]{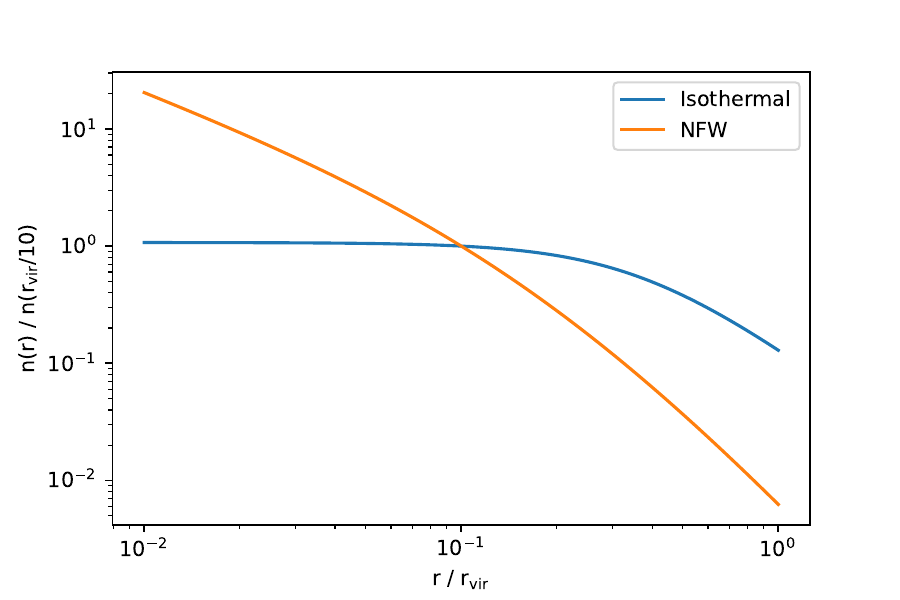}
\caption{The isothermal subhalo number density profile \cite{Diemand_2004} in blue and the NFW profile \cite{nfw_1997} in orange, both normalised to 1 at $r/r_{\mathrm{vir}} = 0.1$. For the NFW profile we set the concentration parameter to $c=1$.}
\label{fig:density_profiles}
\end{figure}
The distribution of the spatial positions of the surviving subhalos within their hosts needs to be specified. To do so, we distribute the subhalos by drawing the distance from the centre of the host halo from a number density profile and drawing their angles about the centre randomly.\\
We considered two profiles. The first one is the NFW profile by \cite{nfw_1997}. It is a standard description of the mass distribution within dark matter halos and has the following number density profile
\begin{equation}
n(r) \propto \frac{1}{(r/r_s)(1 +  r/r_s)^2} ~,
\end{equation}
where $r_s$ is the scale radius, related to the virial radius of the host via the concentration parameter
\begin{equation}
    c = \frac{r_{\mathrm{vir}}}{r_s} ~.
\end{equation}
We compute the concentration parameter from the host halo mass and the redshift using equation (5) in \cite{Vinas_2011} and use the package NFWProfile from the Python package \texttt{halotools}, which is directly based on \cite{nfw_1996}, to sample from the NFW profile.\\
The second one is an isothermal profile described by \cite{Diemand_2004}. It is specifically a fit for the number distribution of substructures within halos. The distribution for the distance $r$ of the subhalo from the centre of its host halo with a virial radius $r_{\mathrm{vir}}$ is given by:
\begin{equation}
n(r) \propto \frac{1}{1 + (r/0.37 r_{\mathrm{vir}})^2}
\end{equation}
Since we use this profile for random sampling, we do not need the normalization of these profiles. We derive the cumulative distribution function by performing the spherical integral and use inverse transform sampling.\\
In figure \ref{fig:density_profiles} we show a comparison of the two profiles, with a normalization to give $n(r)=1$ at $r/r_{\mathrm{vir}} = 0.1$. For the NFW profile we set $c=5$ for this figure.\\
We therefore do not include any environment effects for the clustering of subhalos. The number density profiles depend solely on the mass and redshift of the host. Given that the subhalos are gravitationally bound within their host, we assign them the same redshift. Furthermore, we only draw radii up to the virial radius of the host halo, since the accretion processes simulated in \texttt{PINOCCHIO} do not include substructure that leaves the host halo again.\\
As explained in section \ref{sec:method_merger_tree}, PINOCCHIO does not compute or save any information on the orbit of substructure at the start or during merger. Including an estimation of the orbital velocity of subhalos could therefore not be based on the information provided by PINOCCHIO and would have to be fully drawn from some distribution. Hence, we do not give such an estimate for our resulting halo-subhalo catalog.

\subsection{Comparison with N-body simulations}
\label{sec:method_nbody}
To test the statistical and spatial distribution of our derived subhalos, we compared our halo-subhalo catalogue to the results of a halo finder run on a full N-body simulation. For the latter, we used the halo finder \texttt{ROCKSTAR} \cite{behroozi_2012}, which gets both halos and subhalos and distinguishes between them. As discussed in \cite{2011_knebe}, \texttt{ROCKSTAR} works in 7 dimensions, the 6 phase-space dimensions and time. It is based on an adapted Friends-of-Friends algorithm and is especially useful for studying the evolution, substructure, and merger history of halos. \texttt{ROCKSTAR} also allows to ascertain the corresponding host halo for each subhalo. It therefore gives all the information that we need for the comparison with our halo-subhalo catalogue based on \texttt{PINOCCHIO}.\\ 
We ran \texttt{ROCKSTAR} on snapshots from \texttt{GADGET-2} \cite{gadget2001, gadget2005}, since this N-body code is well tested and can be run on computers with only CPUs. As \texttt{PINOCCHIO} internally uses initial conditions generated with \texttt{N-GenIC}, we used \texttt{N-GenIC} for the initial conditions for \texttt{GADGET-2} with the same seed and the same box size of 200 Mpc/h. With 1024$^3$ simulation particles and at least 10 particles per halo, this gives a mass resolution of $5.7 \cdot 10^9$ M$_\odot$/h. The mass of one simulation particle is about $5.7 \cdot 10^8$ M$_\odot$/h. For these simulations, including the initial condition generator and \texttt{ROCKSTAR} or our subhalo code, respectively, our halo-subhalo catalogue took about 700 times less CPU time to generate than with \texttt{GADGET-2} and \texttt{ROCKSTAR}.\\
In order to estimate the statistical uncertainty of the subhalo properties, we have multiple \texttt{PINOCCHIO} runs with different initial seeds with a box size of 200 Mpc/h and the mentioned resolution. This allows us to calculate mean functions from all these runs and estimate the statistical scatter. Both for \texttt{PINOCCHIO} and \texttt{GADGET-2} we additionally have a few simulations at higher resolution, to study convergence effects. Due to computational limitations they have the same number of simulation particles but smaller box sizes. Full N-body simulations are computationally expensive, therefore we could not run many \texttt{GADGET-2} simulations at the same resolution. While systematic uncertainties of numerical simulations are hard to analize and quantify (see e.g. \cite{2022_massfunctioncalibration}), the statistical uncertainties of \texttt{GADGET-2} and \texttt{ROCKSTAR} are well known.\\
Although having more particles per halo yields more stable mass functions at the low mass end, we can work with few particles per halo for \texttt{PINOCCHIO}, since we do not need to resolve any substructure of the smallest halos here. Future applications like a SHAM model may require more than 10 particles per halo and subhalo, to increase the stability. We have performed convergence studies with \texttt{PINOCCHIO} simulations of different resolutions and found that the subhalo velocity function is stable above 30 particles per halo and subhalo.\\
\texttt{GADGET-2} is a Tree-PM N-body code, resolving gravitational forces down to a softening scale. So the \texttt{GADGET-2} simulations fully reproduce the highly non-linear regime of gravitational clustering, including the hierarchical formation of halos and subhalos, that are then extracted from the distribution of particles with the \texttt{ROCKSTAR} halo finder. However, a numerically stable simulation of subhalos requires a very high resolution (e.g. \cite{2018_bosch, 2018_bosch_2, 2019_ogiya}), so we expect that subhalos found at the resolution we use in cosmological simulations for cosmology are affected by numerical issues. In addition, we ran \texttt{ROCKSTAR} on single snapshots that do not contain information on the past history of particles; working with a full merger tree would help in limiting numerical issues due to limited sampling of halos with few particles.\\
Conversely, the semi-analytic \texttt{PINOCCHIO} code is not affected by these numerical issues, so its approximate validity can be pushed to lower subhalo masses, and its stability with resolution is expected to be better. \texttt{PINOCCHIO} provides halo masses at merger times, but, as already mentioned in Section \ref{sec:method_survival_time}, these masses are not comparable to subhalo masses, that are affected by tidal processes and are numerically unstable. In fact, while the \texttt{PINOCCHIO} halo mass function gives a very good fit to the \texttt{GADGET-2} with \texttt{ROCKSTAR} one, we obtained poorer agreement for the subhalo mass functions, using the mass at merger for \texttt{PINOCCHIO} subhalos.\\
We therefore use the $V_{\rm max}$ quantity to compare \texttt{PINOCCHIO} and numerical subhalos, that is more directly related to the halo mass at merger time. This allowed us to look at subhalo velocity functions instead of subhalo mass functions. The subhalo velocity function is defined as the number of subhalos of a given $V_{\mathrm{max}}$ within a host halo, as a function of host halo mass.\\
For the subhalos derived from the merger history of \texttt{PINOCCHIO}, we do not have the maximum velocity, though. As described in \cite{1998_vc}, the circular velocity of a halo or subhalo can be calculated from its mass $M$ using
\begin{equation}
    V_c = (10 ~ G H(z) M )^{1/3} ~,
\end{equation}
where $G$ is the gravitational constant and $H(z)$ is the Hubble parameter. We then calibrated the ratio between $V_{\mathrm{max}}$ and $V_c$ using the halos from \texttt{ROCKSTAR}, for which both the mass and the maximum velocity are well defined and stable quantities. For the remaining evaluation, we used $V_{\mathrm{max}} = 0.99 V_c$, with the factor 0.99 being independent of redshift. This procedure gave us a relation between the subhalo mass at accretion and the maximum velocity for our simulations based on \texttt{PINOCCHIO}.\\
By performing resolution studies on the subhalo velocity function of both our simulations with \texttt{PINOCCHIO} and the fully numerical simulations with \texttt{GADGET-2}, we found that the results are converged for a subhalo $V_{\mathrm{max}}$ > 60 km/s. This is visible from the cutoff at low $V_{\mathrm{max}}$. Additionally, we have calculated $V_{\mathrm{max}}$ corresponding to the mass resolution of the simulation, meaning the mass of a halo with 10 simulation particles. The relation between mass and $V_{\mathrm{max}}$ depends on redshift, so we used the 99 percentile accretion redshift for the highest mass bin that we considered. This gave about 50 km/s, so using subhalos with $V_{\mathrm{max}}$ > 60 km/s is a conservative choice.\\
One has to be careful not to over-interpret the results from the numerical simulation, though. Even for $V_{\mathrm{max}}$ > 60 km/s, the resolution effects are still large, which leads to a systematic uncertainty. We will discuss this in more detail in section \ref{sec:res_shmf}. Our method to extract subhalos from the merger history of \texttt{PINOCCHIO} is more stable and less affected by resolution effects, so we do not expect a perfect agreement with the numerical simulation. Nevertheless, a comparison allowed us to see whether our results are realistic and in agreement with the results from \texttt{ROCKSTAR} on \texttt{GADGET-2}, within the statistical and systematic uncertainty. For the statistical uncertainty, we calculated the standard deviation of the considered statistics based on 10 \texttt{PINOCCHIO} runs with different random seeds.\\
We performed a comparison on snapshot level at a range of redshifts. Beside the subhalo velocity function, we also studied the halo and subhalo clustering, quantified using their two-point correlation functions (2PCF), in section \ref{sec:res_clustering}. For this purpose, we used the Python package \texttt{Corrfunc} \cite{Corrfunc_2019, Corrfunc_2020}, which provides an accurate and fast calculation of correlation functions from catalogues. 

\section{Results}
\label{sec:results}
 We now describe the results from the statistical comparison of our dark matter halo-subhalo catalogue simulated using \texttt{PINOCCHIO} and the merger history, with the halo catalogues derived from \texttt{ROCKSTAR} run on the N-body simulation \texttt{\texttt{GADGET-2}}. A result of that comparison is also a best fit for the subhalo survival time. The cosmological parameters used are $\Omega_{\Lambda} = 0.73$, $\Omega_{m} = 0.27$, $h = 0.7$, $\Omega_{b} = 0.045$, $\sigma_8 = 0.811$, $n = 0.961$ and $w = -1$, which is close to the WMAP Cosmology \cite{2013_wmap}.

\subsection{Subhalo survival time}
\label{sec:res_fit}
We found our best fit for the subhalo survival time by optimizing the fitting parameters to make the subhalo velocity functions agree with the one from the N-body simulation. The number of surviving subhalos is matched by the normalization, while the mass distribution is given by the shape of the subhalo velocity function. For our simulations, there is a direct relation between the subhalo velocity and the subhalo mass.\\
For the reference simulation with \texttt{GADGET-2} and \texttt{ROCKSTAR}, we used a box size of 200 Mpc/h and 1024$^3$ particles, leading to a particle mass of about $5.7 \cdot 10^8$ $M_\odot$/h. We ran \texttt{ROCKSTAR} with a lower limit of 10 particles per halo, but could not use the smallest halos and subhalos, due to limited convergence of the fully numerical simulation.\\
Note that these results are only applicable to our implementation. We have indeed tailored our best fit survival time to work well for \texttt{PINOCCHIO} and its merger history. Our results therefore depend on the approximate nature and the properties of \texttt{PINOCCHIO} as a simulation. Furthermore, our comparison also depends on the halo finder that is used together with the N-body simulations. As mentioned previously, simulating substructure with a fully numerical simulation requires a very high resolution. We have performed convergence studies and included the systematic uncertainty as explained in section \ref{sec:res_shmf}, but there is still some dependence on the resolution.\\
As described in section \ref{sec:method_survival_time}, we used equation (\ref{eqn:merger_time_form}) for the functional form of the subhalo survival time. We found that the parameter $b$ has to be set lower than presented in \cite{boylon-kolchlin2008} or \cite{Jiang_2008}. This is especially the case at higher redshifts, such that we had to include a dependence of $b$ on the observation redshift. The observation redshift in this case refers to the redshift of the considered snapshot or to the redshift of the host halo in the lightcone. We kept the dependence on the orbital circularity $\eta$ and sample $\eta$ as explained in section \ref{sec:method_survival_time}. As described in section \ref{sec:method_survival_time}, we set the parameter $d$ to $d=0$, since we could only draw the orbital energy uniformly. For the other fitting parameters $A$ and $c$, we found that it is possible to use constant values for low enough redshifts. The inclusion of subhalos is especially important for applications at lower redshift since this is where the resolution has to be high. At higher redshift one observes the bright tale of the galaxy luminosity function, that is dominated by central galaxies, so the need of accuracy in populating halos with satellites is less stringent.\\
For higher redshifts $z > 1$ (given our choice of cosmological parameters), lower values of the normalization parameter $A$ are needed for our simulation to be in agreement with the reference simulation. This makes the functional form somewhat more complicated due to more free parameters.\\
For the redshift range $0 \leq z \lesssim 3$, we found the following best fit parameters:
\begin{align}
\begin{split}
A &=\begin{cases}
    0.195 &\mathrm{for} ~ D(z) \geq 0.6\\
    0.195 \cdot \big(\frac{D(z)}{0.6}\big)^{2} &\mathrm{for} ~ D(z) \leq 0.6
    \end{cases}\\
b &= 0.92 \cdot D(z) \\
c &= 1.9 \\
d &= 0.0 
\end{split}
\end{align}
with the normalized linear growth rate factor $D(z)$ as defined in section \ref{sec:method_survival_time}. This leads to values of $b=0.92$ at $D=1$ ($z = 0$) and $b=0.57$ at $D=0.6$ ($z \approx 1$) for our chosen cosmological parameters. The value of $c$ is the same as in \cite{boylon-kolchlin2008}. For $A$, our constant fitting value at low redshifts corresponds to their value including the factor of 0.9 for $A$ due to baryonic bulges. \\
The parameter $A(D)$ can also be written in the following form, where $\Theta(x)$ is the Heaviside step function:
\begin{equation}
    A(D) = 0.195 \cdot \Big[1 + \Theta(0.6 - D) \cdot \Big(\Big(\frac{D}{0.6}\Big)^2 - 1\Big)\Big]
\end{equation}
This finally gives us our best fit for the merger or subhalo survival time:
\begin{equation}
\label{eqn:t_merge}
    \tau_{\mathrm{merge}} = A(D) ~ \tau_{\mathrm{dyn}} ~  \frac{(M_{\mathrm{host}}/m_{\mathrm{sub}})^{0.92 \cdot D}}{\ln(1+M_{\mathrm{host}}/m_{\mathrm{sub}})} \exp \big(1.9 \eta \big)
\end{equation}
The figures displayed in sections \ref{sec:res_shmf} and \ref{sec:res_clustering} are from simulations with \texttt{PINOCCHIO} and the fitting parameters presented above for the subhalo survival time.\\
Although the parameters for the subhalo mass function depend on the chosen cosmological parameters, we expect the fit to still work for similar cosmologies. We have introduced the redshift dependence in $b$ via $D(z)$ to minimize the dependence on cosmology of this fit. It should be considered for a parameter inference, though.\\
Using the subhalo velocity functions, we could only test this fit for the subhalo survival time for subhalos with $V_{\mathrm{max}}$ > 60 km/s due to the limited convergence of the fully numerical simulation we compared to. Nevertheless, we can also apply it to smaller subhalos, using that our simulation based on \texttt{PINOCCHIO} is more stable for lower resolutions.

\begin{figure}[h!]
\centering 
\includegraphics[width=.85\textwidth,angle=0]{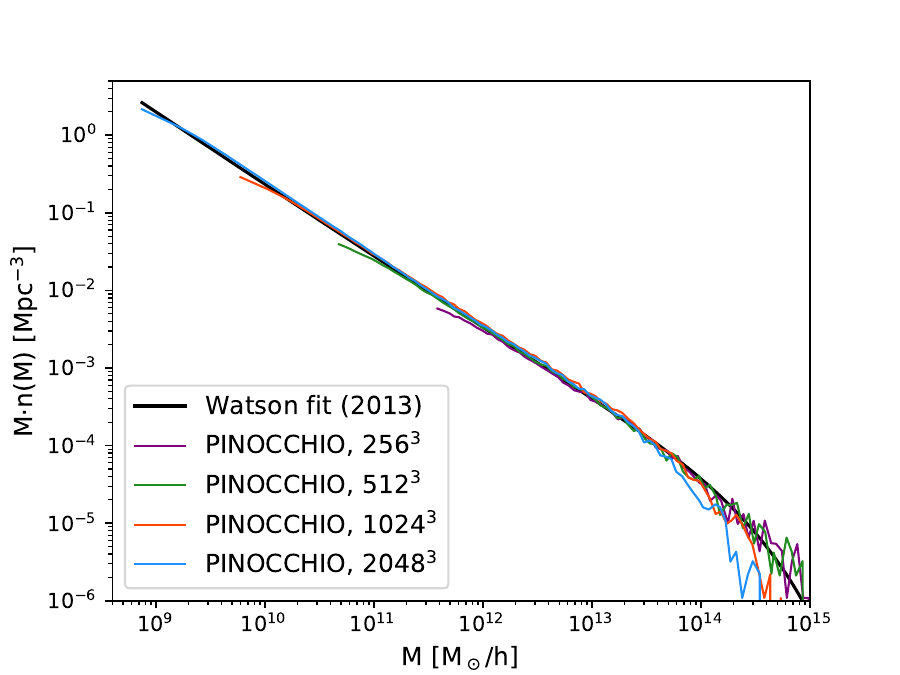}
\caption{Comparison of the halo mass functions for \texttt{PINOCCHIO} with different resolutions and with the fitting function by Watson et al. \cite{2013_Watson}. All simulations have a box size of 200 Mpc/h and the same initial conditions.}
\label{fig:mf_halos}
\end{figure}

\subsection{Subhalo velocity function}
\label{sec:res_shmf}
The accuracy of the halo mass function and the halo power spectrum have been shown and discussed by \cite{pinocchio2013} and \cite{pinocchio2017}. As shown in \cite{pinocchio2013}, the halo mass function agrees well with N-body simulations up to about $10^{14}$ – $10^{15}$ $M_\odot$/h, with the upper limit decreasing with increasing redshift.\\
In figure \ref{fig:mf_halos}, we show a comparison between the halo mass functions of different \texttt{PINOCCHIO} simulations. The same cosmological parameters are used, but the resolution varies. All the simulations have the same box size (200 Mpc/h) and the same random seed, leading to the same large-scale structure realization. As a reference, we also show the fit by \cite{2013_Watson}, which was used in these simulations as the reference mass function. To resolve halos and subhalos hosting single galaxies and dwarf galaxies, we need a high resolution. We notice that \texttt{PINOCCHIO} underestimates the halo mass function at the high mass end, which only appears for very high-resolution simulations or small box sizes. This effect is visible in the mass function of the two simulations with the highest resolution in figure \ref{fig:mf_halos} and was shown and discussed in \cite{pinocchio2013}. This behaviour of the code is under investigation and might be resolved in newer versions of \texttt{PINOCCHIO}.
\begin{figure}[ht]
\centering 
\includegraphics[width=0.95\textwidth,angle=0]{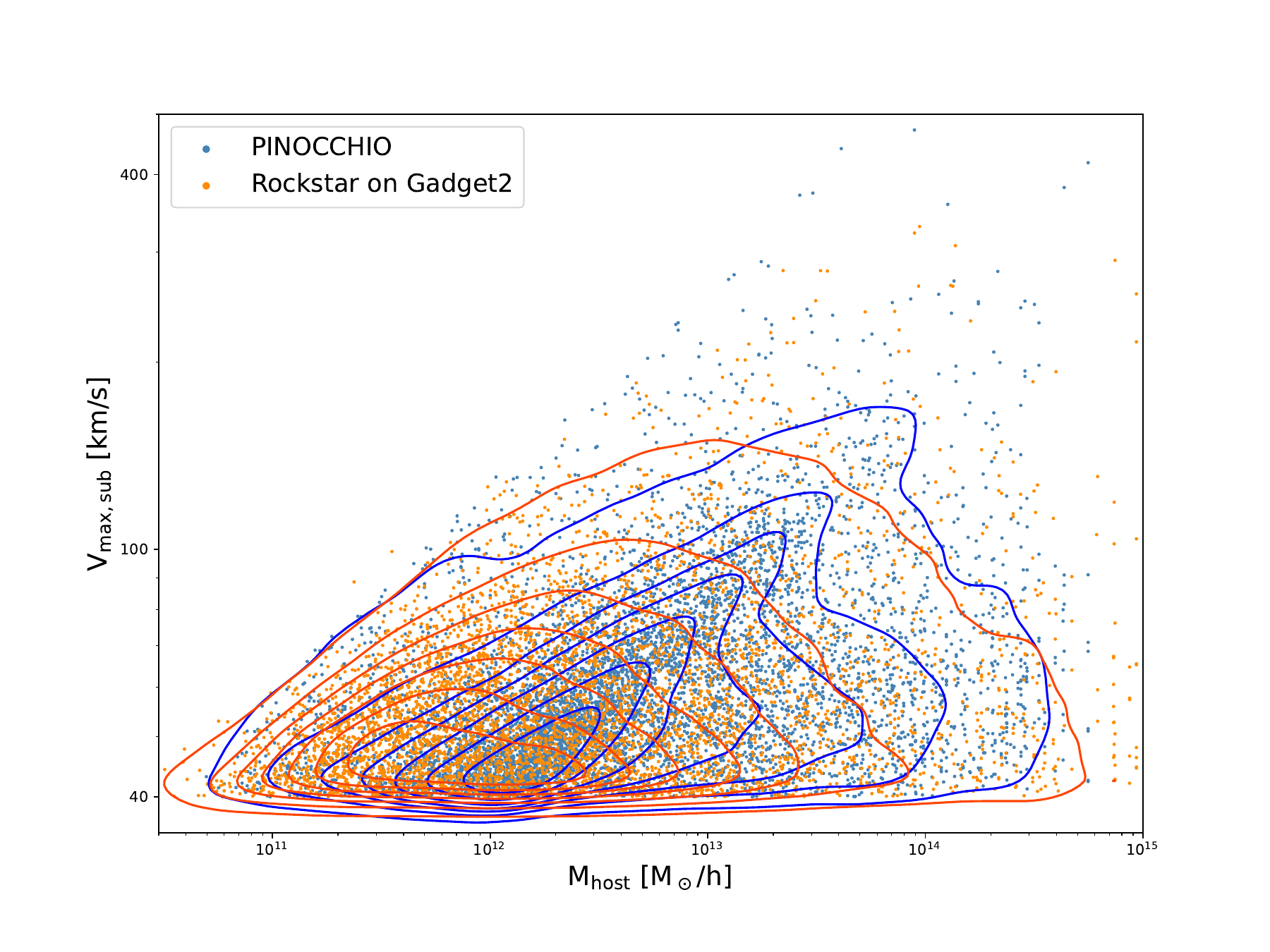}
\caption{Host halo mass ($M_{\mathrm{host}}$) to subhalo maximum velocity ($V_{\rm max, sub}$, at accretion) distribution at redshift $z=0$ for our subhalos in \texttt{PINOCCHIO} (blue / lightblue) and \texttt{ROCKSTAR} run on  \texttt{GADGET-2} (red / orange). Both simulations have a box size of 200 Mpc/h and $1024^3$ particles. Only subhalos with $V_{\rm max, sub}$ > 40 km/s are shown. For clarity, a random selection of 5000 subhalos is plotted for each simulation. The contours were calculated using the full catalogs.}
\label{fig:m_vmax_relation}
\end{figure}
\begin{figure}[t]
\centering 
\includegraphics[width=0.88\textwidth,angle=0]{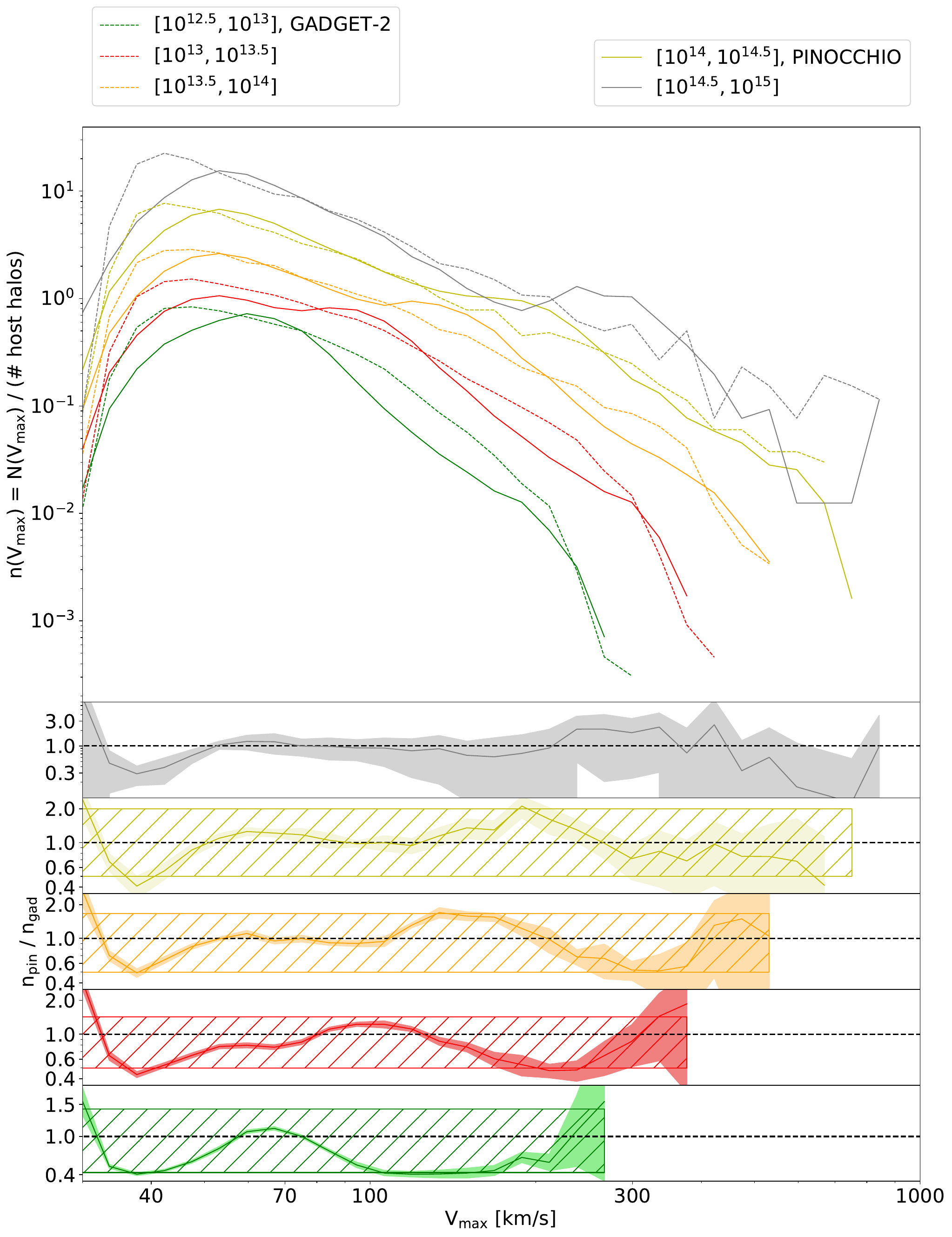}
\caption{Top panel: Subhalo velocity function, defined as the number of subhalos of velocity V$_{\rm max, sub}$ per host halo, at redshift $z=0$. Different colours refer to different host halo mass bins (in units of M$_\odot$/h), solid lines are for our subhalo code with \texttt{PINOCCHIO} averaged over 10 runs, dashed lines are for \texttt{ROCKSTAR} run on \texttt{GADGET-2}. All simulations have a box size of 200 Mpc/h and $1024^3$ simulation particles. Lower panels: Ratio between the subhalo velocity functions of \texttt{PINOCCHIO} and \texttt{GADGET-2}. The shaded areas show the statistical error, corresponding to 2$\sigma$ based on 10 runs for \texttt{PINOCCHIO}. The hatched areas display the systematic uncertainties of \texttt{GADGET-2} with \texttt{ROCKSTAR}.}
\label{fig:shmf_z0}
\end{figure}
\begin{figure}[ht!]
\centering 
\includegraphics[width=0.7\textwidth,angle=0]{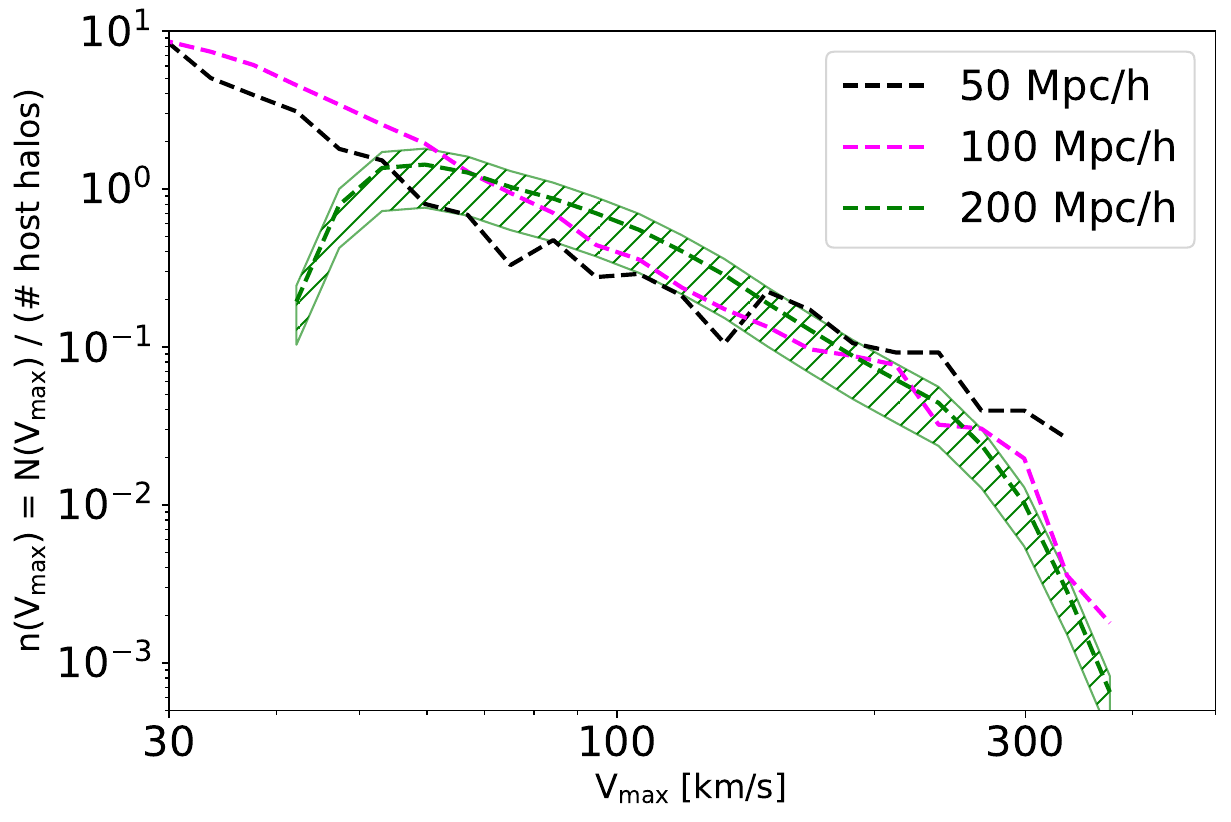}
\caption{Subhalo velocity functions at redshift $z=1$ for the host halo mass bin $[10^{12.5}, 10^{13}]$ M$_\odot$/h, for \texttt{GADGET-2} with \texttt{ROCKSTAR}. The different colors refer to different simulation box sizes. The number of simulation particles is always 1024$^3$. The hatched area shows our estimate for the systematic uncertainty of the simulation with the lowest resolution.}
\label{fig:convergence}
\end{figure}\\
We checked the retrieval of subhalos with our merger time estimation code by considering subhalo statistics. Figure \ref{fig:m_vmax_relation} shows the relation between the maximum velocity $V_{\mathrm{max}}$ of a subhalo and the mass of its host. As described in section \ref{sec:method_nbody}, $V_{\rm max, sub}$ is calculated from the subhalo mass at accretion for our simulations. The host halo mass is defined as its total mass (including the mass of its subhalos) at the observed redshift. The relation agrees approximately with the one for \texttt{ROCKSTAR} applied to \texttt{GADGET-2}. We see that \texttt{ROCKSTAR} includes host halos at lower masses for the same $V_{\rm max, sub}$. The difference at the high mass end of host halos is due to \texttt{PINOCCHIO} not providing enough such large halos, as described earlier in this section and in figure \ref{fig:mf_halos}.\\ 
In figures \ref{fig:shmf_z0} and \ref{fig:shmf_z1} we display the subhalo velocity function, i.e. the average number of subhalos as a function of subhalo maximum velocity $V_{\rm max}$ residing in a host halo of a given mass. We have performed this comparison at multiple redshifts, but are only showing the results of $z=0$ and $z=1$ here. Figure \ref{fig:shmf_z0} is for redshift $z=0$, while figure \ref{fig:shmf_z1} is for $z=1$. The solid (dashed) lines in the upper panels refer to our \texttt{PINOCCHIO} (\texttt{ROCKSTAR} on \texttt{GADGET-2}) subhalos, and the different colours refer to different host halo mass bins (in M$_\odot$/h). Note that \texttt{ROCKSTAR} returns subhalos with fewer particles per object, which are excluded in our code due to the minimum number of particles per group that we set to 10, both for halos and subhalos. Since the subhalos from \texttt{ROCKSTAR} with less than 30 particles are unstable and gave erratic values for $V_{\rm max}$, we additionally excluded those. We see that both the normalization and the dependence on host halo mass and $V_{\rm max, sub}$ agree well. These subhalo mass functions are created from one single simulation run of \texttt{GADGET-2} and 10 runs of \texttt{PINOCCHIO}, with a box size of 200 Mpc/h and 1024$^3$ simulation particles.  As explained in section \ref{sec:method_nbody}, we do not have a full set of \texttt{GADGET-2} simulations at the same resolution, therefore we are showing the result of one simulation and don't show an estimate of the statistical uncertainty. While the systematic uncertainty is very different, the statistical scatter is similar to that of our simulations based on \texttt{PINOCCHIO}.\\
For a more thorough comparison, we additionally show the ratio of the subhalo velocity functions from the different simulations in the lower panels of figures \ref{fig:shmf_z0} and \ref{fig:shmf_z1}. The shaded areas correspond to the 2$\sigma$ statistical uncertainty estimated from 10 runs of \texttt{PINOCCHIO}. The statistical uncertainty is larger for higher mass bins of the host halos and for larger $V_{\rm max, sub}$.\\
The hatched areas display the systematic uncertainty of \texttt{GADGET-2} with \texttt{ROCKSTAR}. We estimated this uncertainty by running \texttt{GADGET-2} at different much higher resolutions, using smaller simulation boxes. All simulations have 1024$^3$ simulation particles, with a box size varying between 50 and 200 Mpc/h.\\
In figure \ref{fig:convergence}, we show the subhalo velocity function for \texttt{GADGET-2} simulations with different resolutions at redshift $z=1$, for the mass bin $[10^{12.5}, 10^{13}]$ M$_\odot$/h. The resolution dependence is large enough to give significant systematic uncertainty. Given the limited statistical stability of a very small box like 50 Mpc/h, we mainly use the difference between the boxes with 100 and 200 Mpc/h to estimate the systematic uncertainty. For each considered redshift, we calculated the relative difference between the subhalo velocity function from Rockstar run with GADGET-2 at different resolutions. This allows us to estimate a range in which the velocity function should lie. Since there are fewer high mass halos in small boxes, this method does not allow us to estimate the systematic uncertainty of the highest mass bin of host halos at redshift $z=0$.\\
For substructure to be simulated in a stable manner with a fully numerical simulation, the resolution must be high. It is therefore not surprising that the subhalo velocity function is not fully converged and that the systematic uncertainty is rather large. Numerical Simulations already have resolution effects for halo properties, as discussed by e.g. \cite{2022_massfunctioncalibration}. Nonetheless, it is noteworthy how large the systematic uncertainties are here. While a halo can be simulated in a stable manner with about 30 simulation particles, subhalos require far more for their statistics to be fully converged. Since subhalos are mostly small structures, this means that the resolution of the simulation needs to be even higher. For our simulations based on \texttt{PINOCCHIO}, the resolution effect is less severe. We have done the same comparison as shown in figure \ref{fig:convergence} also for our simulations and found far less deviations.
\begin{figure}[ht!]
\centering 
\includegraphics[width=0.98\textwidth,angle=0]{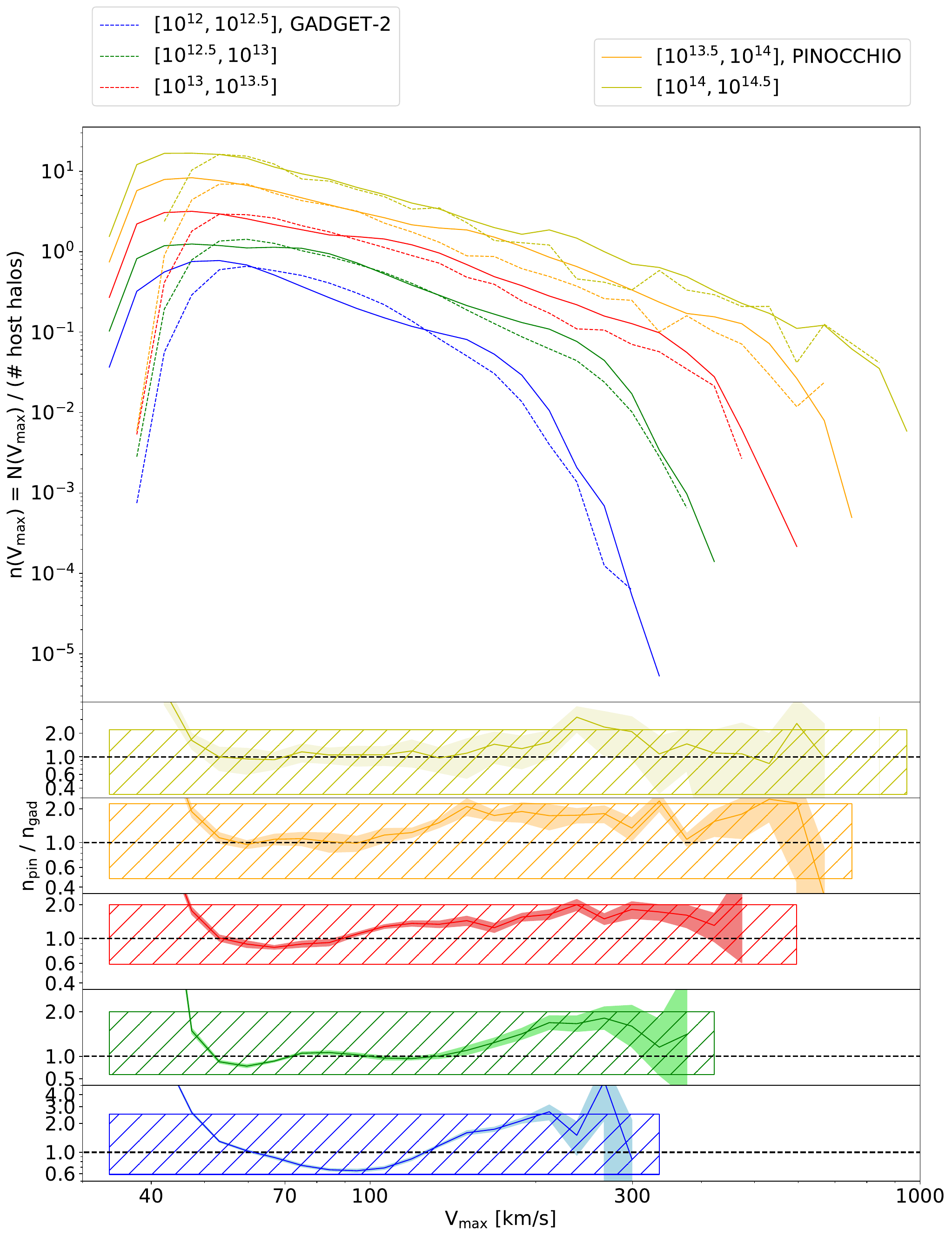}
\caption{Same as figure \ref{fig:shmf_z0}, but at redshift $z=1$. There are fewer high mass halos at higher redshifts. The shaded areas in the lower panels show the statistical uncertainty of \texttt{PINOCCHIO}, the hatched areas the systematic uncertainty of \texttt{GADGET-2} with \texttt{ROCKSTAR}.}
\label{fig:shmf_z1}
\end{figure}\\
\noindent The shaded areas in the lower panels in figure \ref{fig:shmf_z1} show the statistical uncertainty of our simulations and the systematic uncertainty of \texttt{GADGET-2} with \texttt{ROCKSTAR} is shown as hatched areas. This is analogous to figure \ref{fig:shmf_z0}, only at redshift $z=1$. The hatched area in figure \ref{fig:convergence} corresponds to the green hatched area in figure \ref{fig:shmf_z1}, which displays the systematic uncertainty of the same mass bin.\\
These figures are made using the best-fit subhalo survival time presented in section \ref{sec:res_fit}. As mentioned in section \ref{sec:method_nbody}, a comparison is only possible for $V_{\rm max} >$ 60 km/s. The highest mass bin agrees due to the statistical uncertainty. The remaining mass bins are in agreement due to the systematic uncertainty of the numerical reference simulation. The agreement is at the same level for the full redshift range $0 \leq z \lesssim 3$ of our fit for the subhalo survival time. Although the limited convergence and the resulting systematic uncertainty do not allow for a more precise calibration, this comparison shows that our method gives a realistic halo-subhalo catalogue.

\subsection{Halo and subhalo clustering}
\label{sec:res_clustering}
The spatial distribution of the simulated halos and subhalos is important, since this determines the clustering of galaxies after a subhalo abundance matching, in addition to the number statistics and the galaxies' assignment to halos.\\
The simulated halos are clustered according to the procedure in \texttt{PINOCCHIO} and their clustering properties have been well tested. In \cite{pinocchio2017} it is shown (in figure 6) that the halo power spectrum of \texttt{PINOCCHIO} agrees well with that from an N-body simulation for around $k < 0.5$. Above that, meaning on even smaller scales, the power spectrum tends to be underestimated by \texttt{PINOCCHIO}. By consequence, the two-point correlation function of halos is underestimated on small scales. Small-scale clustering is predicted better by \texttt{PINOCCHIO} if the mass cut for halos is at a larger number of simulation particles per halo, as shown in \cite{pinocchio2017}. Note that we set the minimum number of particles per halo to a relatively low value in order to simulate sufficiently small halos in a large enough volume.\\
While \texttt{PINOCCHIO} does not track substructure after a merger and therefore does not resolve subhalos, we now have subhalos from the merger tree. The correlation function for the subhalos alone and therefore the total halo and subhalo clustering depends on how the subhalos are distributed within their hosts. Since we do not have spatial resolution within the halos, we place the surviving subhalos according to a distribution that we sample. In our case, this is given by a random angular distribution and a density profile for the radial distance from the centre of the host. The simulations used for the figures in this section have a box size of 200 Mpc/h and 1024$^3$ simulation particles, unless stated otherwise. We use \texttt{Corrfunc} \cite{Corrfunc_2019, Corrfunc_2020} to calculate all the halo and subhalo two-point correlation functions (2PCF) shown for both simulations. Note that we use the subhalo mass from \texttt{ROCKSTAR} in this section. Since we consider large mass bins, the inaccuracy of the subhalo masses is less of a problem than for a subhalo mass function, for example. Nevertheless, the systematic uncertainty is not negligible, as we will discuss later.
\begin{figure}[ht]
\centering 
\includegraphics[width=0.85\textwidth,angle=0]{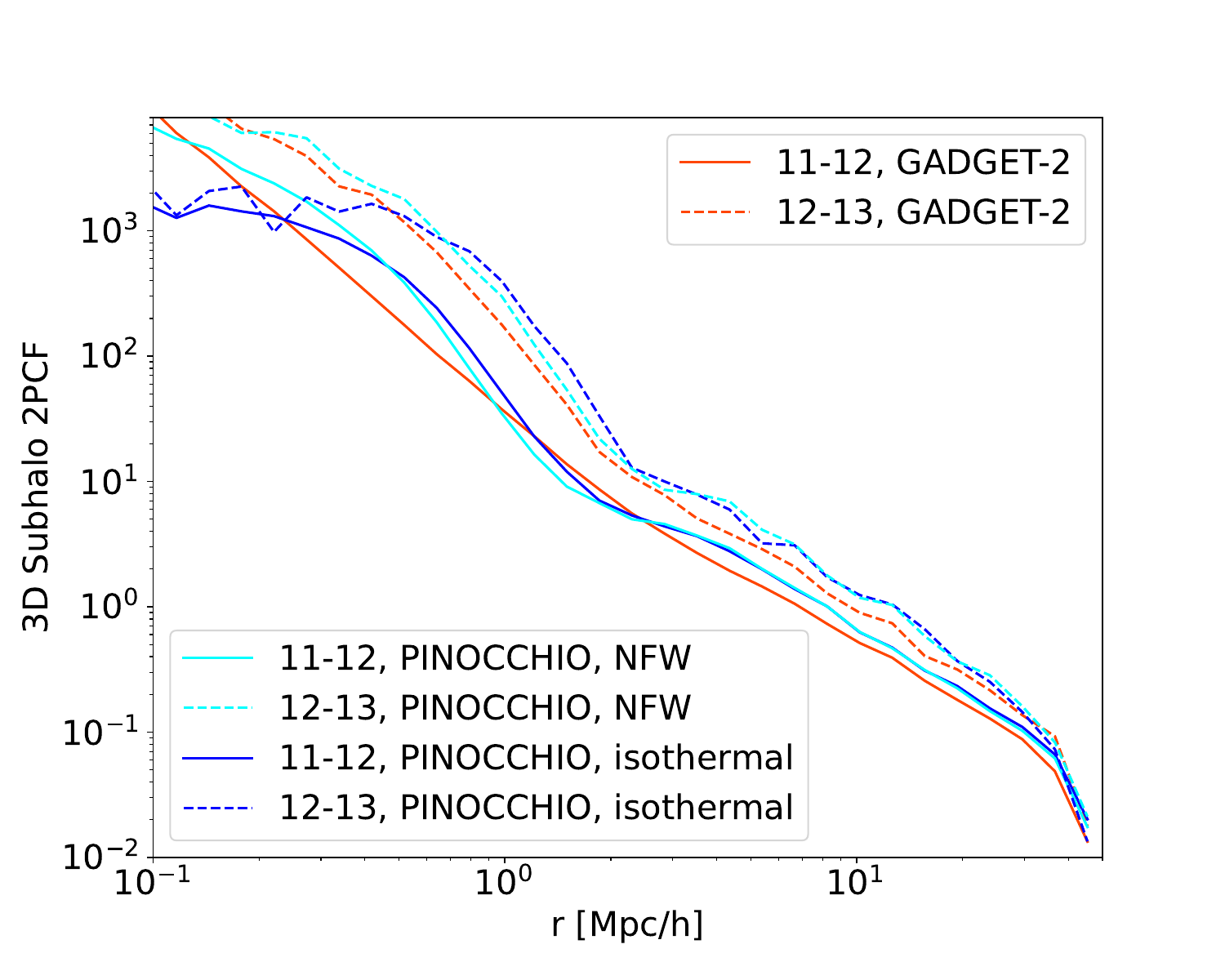}
\caption{Subhalo 3D two-point correlation function at $z=0$ for our code with \texttt{PINOCCHIO} (blue and cyan) and \texttt{ROCKSTAR} on \texttt{GADGET-2} (red), calculated with \texttt{Corrfunc}. The mass bins $[10^{11}, 10^{12}]$ M$_\odot$/h and $[10^{12}, 10^{13}]$ M$_\odot$/h are shown. The cyan lines for \texttt{PINOCCHIO} are with an NFW profile for the distribution of subhalos inside the hosts, the blue lines are for an isothermal profile.}
\label{fig:subhalo_2pcf}
\end{figure}
\begin{figure}[ht!]
\centering 
\includegraphics[width=0.95\textwidth,angle=0]{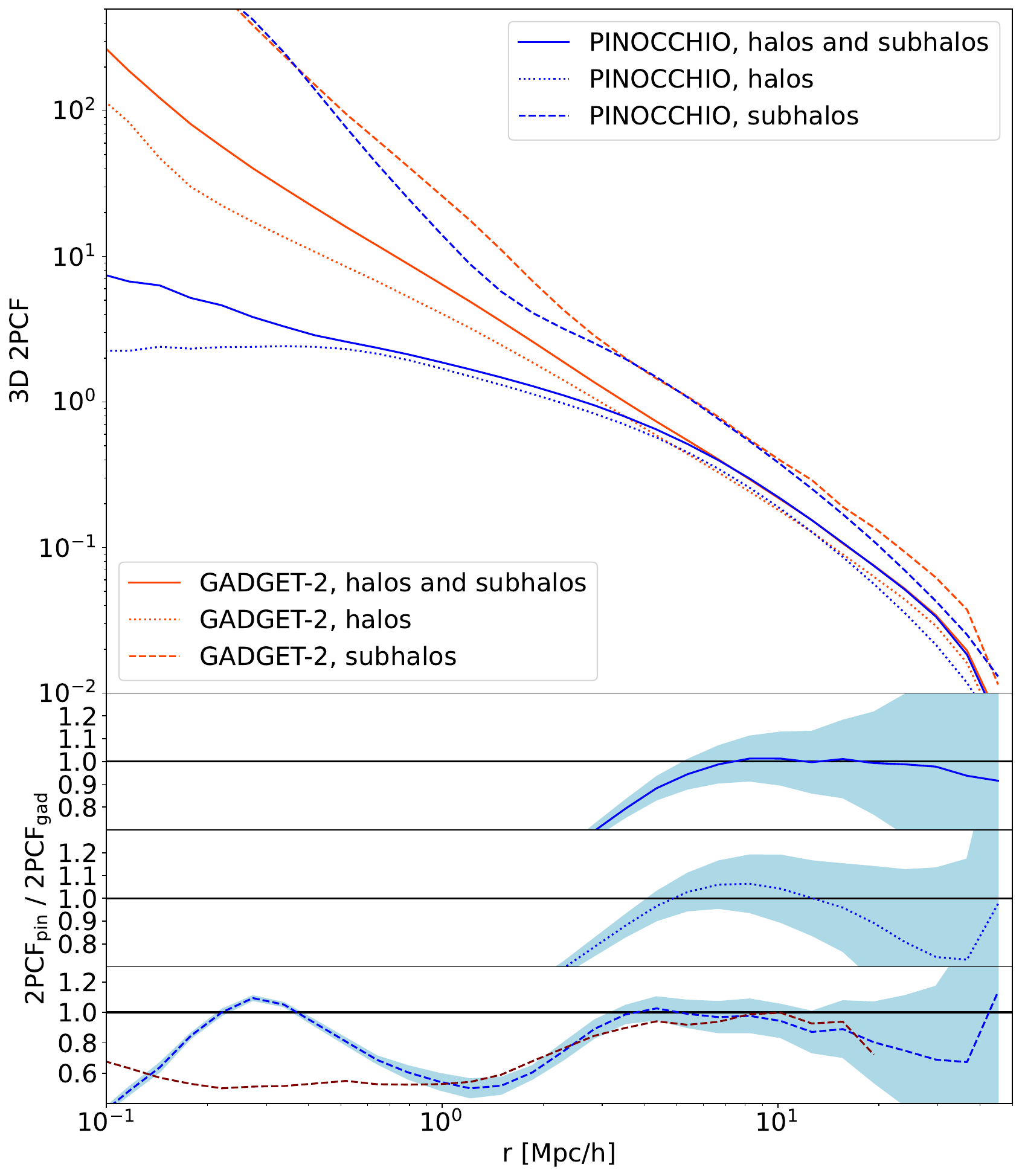}
\caption{Top panel: Halo and subhalo 3D two-point correlation function at $z=0$ for our code with \texttt{PINOCCHIO} (blue) and \texttt{ROCKSTAR} on \texttt{GADGET-2} (red), calculated with \texttt{Corrfunc}. Only the mass bin $[10^{10}, 10^{11}]$ M$_\odot$/h is shown. The solid lines correspond to the 2PCF using both halos and subhalos, the dotted lines only halos and the dashed lines only subhalos. For \texttt{PINOCCHIO}, the mean results from 10 runs are shown. Lower panels: Ratio between the 2PCF of \texttt{PINOCCHIO} and \texttt{GADGET-2}. The shaded areas show the statistical error, corresponding to 2$\sigma$ based on 10 runs for \texttt{PINOCCHIO}. For the subhalos, we additionally show the ratio between a higher resolution \texttt{GADGET-2} simulation and the reference simulation in dark red, to show the systematic uncertainty of \texttt{ROCKSTAR} on \texttt{GADGET-2} for subhalos.}
\label{fig:halo_2pcf_separate}
\end{figure}\\
\noindent In figure \ref{fig:subhalo_2pcf}, we show the 3D two-point correlation function of subhalos only at redshift $z=0$, for the two different distribution functions presented in section \ref{sec:method_distribution}. Two subhalo mass bins are shown, $[10^{11}, 10^{12}]$ and $[10^{12}, 10^{13}]$ M$_\odot$/h, to show the dependence on mass. We see that the difference between the isothermal profile and the NFW profile introduced in section \ref{sec:method_distribution} is small at separations above 0.3 Mpc/h. The behaviour at around 4 Mpc/h, including the kink, is due to the transition from the 1-halo term to the 2-halo term dominating. At large separations the 2-halo term dominates, therefore the two density profiles give the same two-point correlation function in this regime. Both profiles result in a slightly higher correlation function compared to \texttt{ROCKSTAR} with \texttt{GADGET-2}. This discrepancy comes from the different definitions of subhalos in the halo finder and our subhalo model. For the remaining comparison and the results shown in figures \ref{fig:halo_2pcf_separate} and \ref{fig:halo_2pcf}, we chose to use the isothermal profile, which was specifically fitted by \cite{Diemand_2004} for the subhalo distribution within halos.\\
In the upper panel of figure \ref{fig:halo_2pcf_separate}, we show the 3D two-point correlation function for the mass bin $[10^{11}, 10^{12}]$ M$_\odot$/h at redshift $z=0$, treating halos and subhalos separately. While we use only one \texttt{GADGET-2} run (red lines), the correlation functions from our simulations based on \texttt{PINOCCHIO} (blue lines) are the means from 10 simulation runs with different initial conditions. As described before, the halo clustering from \texttt{PINOCCHIO} is underestimated on small scales. This is visible in the dotted blue line of figure \ref{fig:halo_2pcf_separate}. In our simulation where we push the halo mass limit to small halos, we see that the halo two-point correlation function is low for $r < 1$ Mpc/h. The underestimation at small scales is reduced a lot by including subhalos, but is still noticeable especially for $r < 0.3$ Mpc/h. The total correlation function also depends on the number of surviving subhalos, being higher for a larger subhalo number ratio.\\
In the lower panels of figure \ref{fig:halo_2pcf_separate}, the ratios between the respective two-point correlations functions from our simulations and the reference simulation are shown. The shaded areas are the statistical 2$\sigma$ uncertainties estimated from the 10 \texttt{PINOCCHIO} runs. On scales above about 5 Mpc/h, our simulations are in statistical agreement with the reference simulation. On smaller scales, the halo two-point correlation function and the total two-point correlation function are not in agreement due to the underestimation of halo clustering on small scales by \texttt{PINOCCHIO}. In the lowest panel, we additionally show the ratio between the subhalo two-point correlation function of a higher resolution \texttt{GADGET-2} run and the reference simulation. One simulation particle in this simulation of higher resolution has a mass of about $7.1 \cdot 10^7$ M$_\odot$/h. We can see that the systematic uncertainty of subhalo clustering on small scales is large enough for the numerical simulation to make our simulations in agreement.\\
Although the contribution of subhalos to the overall correlation function is especially important on small scales corresponding to the 1-halo regime, subhalos also have a 2-halo term. Given that each subhalo is hosted by a halo of a mass higher than its own, and massive halos are more clustered than small ones, the 2-halo term of the power spectrum and therefore of the correlation function is higher for subhalos than for halos in the same mass bin. There are far more halos than subhalos in each mass bin, though, so they do not dominate the overall correlation function.
\begin{figure}[b]
\centering 
\includegraphics[width=0.8\textwidth,angle=0]{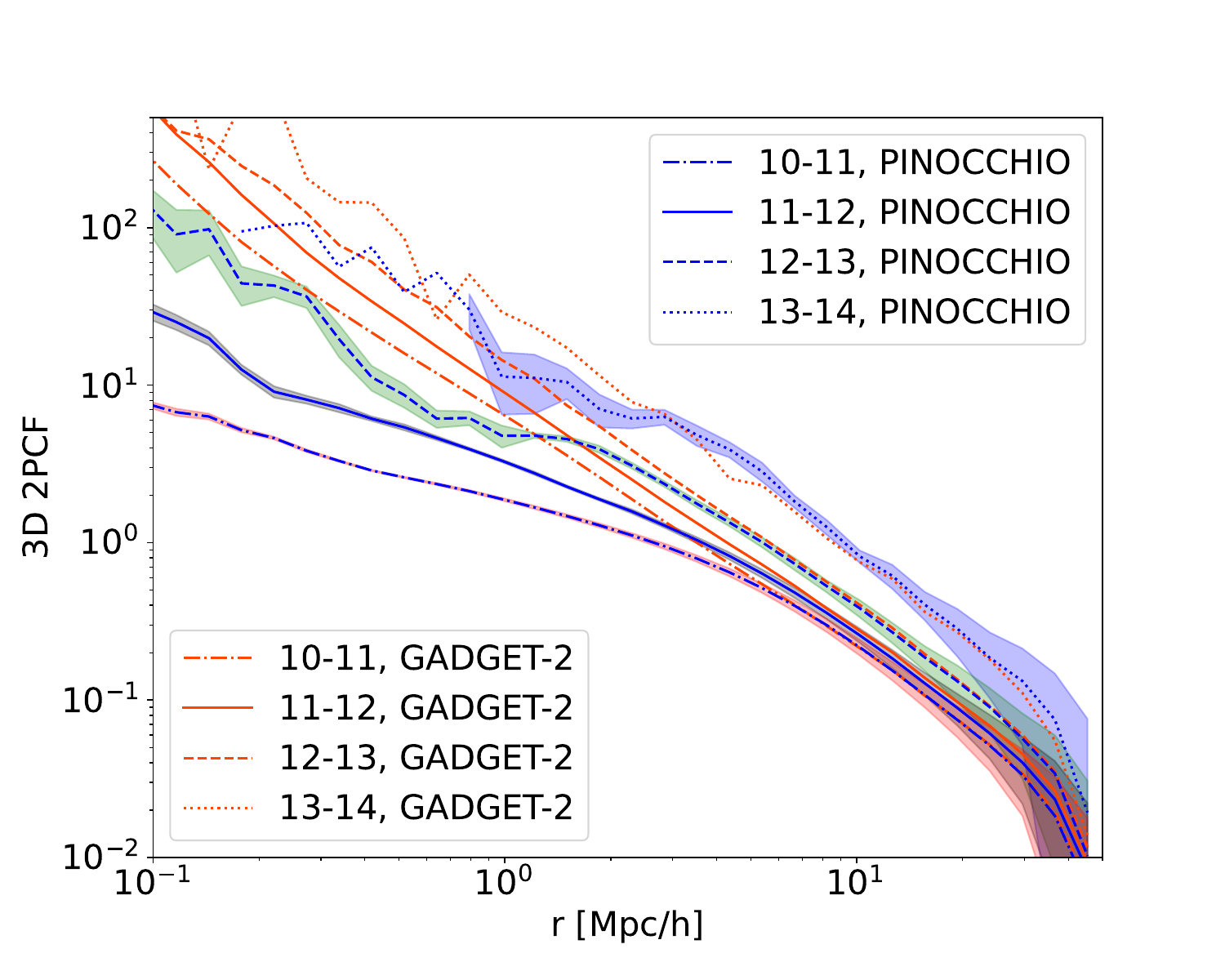}
\caption{3D two-point correlation function at $z=0$ of halos and subhalos combined, for our code with \texttt{PINOCCHIO} (blue) and \texttt{ROCKSTAR} on \texttt{GADGET-2} (red), calculated with \texttt{Corrfunc}. Dot-dashed, solid, dashed and dotted lines refer to the mass bins $[10^{10}, 10^{11}]$, $[10^{11}, 10^{12}]$, $[10^{12}, 10^{13}]$ and $[10^{13}, 10^{14}]$ M$_\odot$/h, respectively. The shaded areas correspond to the statistical 2$\sigma$ uncertainty calculated from 10 \texttt{PINOCCHIO} runs.}
\label{fig:halo_2pcf}
\end{figure}\\
\noindent Figure \ref{fig:halo_2pcf} shows the halo-subhalo two-point correlation function in three dimensions at redshift $z=0$, for different mass bins. Both halos and subhalos are considered, without distinguishing between them. Red vs. blue lines correspond to \texttt{PINOCCHIO} vs. \texttt{ROCKSTAR} applied to \texttt{GADGET-2}, while different line styles correspond to different mass bins. The mass bins are described by $\log_{10}$ of the halo mass in M$_\odot$/h in the legend. The highest mass bin is less stable due to weak statistics. The two-point correlation function for our simulations are again the means from 10 simulation runs. The shaded areas represent the 2$\sigma$ statistical uncertainty estimated from these 10 runs. We can see that the two simulations are in statistical agreement on medium to large scales above about 10 Mpc/h.\\
The main source of statistical uncertainties, especially on large scales, is the limited box size. While having larger simulation boxes would be favourable to reduce the statistical uncertainty, it makes having enough resolution more difficult. For our code based on \texttt{PINOCCHIO} we are able to run simulations at the same resolution with twice the box size. For \texttt{GADGET-2} we are not able to run simulations with much more particles, therefore we could not do the comparison for 400 Mpc/h boxes. We expect similar statistical uncertainties for \texttt{ROCKSTAR} on \texttt{GADGET-2}, but don't have enough simulation runs at the same resolution to stably quantify them. For halo properties there have been studies in the past comparing the statistical scatter of \texttt{PINOCCHIO} to that of N-body simulations.\\
Our resulting halo and subhalo correlation functions show some remaining differences compared to the N-body simulation, as visible in figures \ref{fig:subhalo_2pcf}, \ref{fig:halo_2pcf_separate} and \ref{fig:halo_2pcf}. The differences are mainly at small scales. At larger scales, the differences are within the statistical uncertainties. The same level of agreement holds for higher redshifts. When galaxies are added to the halos and subhalos with an HOD or a SHAM approach, the model is calibrated by comparing to a data set. This calibration can absorb small differences in the halo/subhalo two-point correlation function.

\section{Conclusions}
\label{sec:conc}
In this paper, we have presented our forward model for the fast simulation of catalogues of dark matter halos and subhalos. We described our method in section \ref{sec:method}. The main ingredients are as follows:
\begin{itemize}
    \setlength\itemsep{0.2em}
    \item Halo catalogue in snapshot or lightcone format and merger history from \texttt{PINOCCHIO};
    \item Extraction of the merging subhalos for all the halos using the merger history;
    \item Determination of the surviving subhalos by estimating the merger time corresponding to the subhalo survival time;
    \item Spatial distribution of the subhalos within their hosts using a number density profile, assigning them the mass of the subgroup at merger.
\end{itemize}
We introduced a new fitting function for the subhalo survival time, depending on the mass ratio of host and subhalo mass at merger. The dependence on $M_{\mathrm{host}}/m_{\mathrm{sub}}$ we used is of the same functional form as presented in e.g. \cite{Jiang_2008} and \cite{boylon-kolchlin2008}. In section \ref{sec:res_fit} we show our best fit for the subhalo survival time for \texttt{PINOCCHIO} subhalos extracted from the merger history, which is given by equation (\ref{eqn:t_merge}).\\
In section \ref{sec:res_shmf}, we compared the relation between host mass and subhalo $V_{\mathrm{max}}$ and the subhalo velocity function of our model with the results for the N-body simulation \texttt{GADGET-2} with the halo finder \texttt{ROCKSTAR}. For this comparison, we worked on snapshot level at different redshifts with a resolution leading to halos with masses above $5.7 \cdot 10^9$ M$_\odot$/h for both simulations. For a stable comparison, we only considered subhalos with $V_{\mathrm{max}}$ > 60 km/s. We have calculated the statistical uncertainty using multiple \texttt{PINOCCHIO} runs with different initial conditions. To estimate the systematic uncertainty of the fully numerical reference simulation, we have compared the results from simulations with different resolutions. We find that the subhalo velocity functions are in agreement on the relevant mass scales within the given uncertainties.\\
We presented a comparison for the clustering of the halos and subhalos with the same \texttt{GADGET-2} with \texttt{ROCKSTAR} simulation in section \ref{sec:res_clustering}. We used \texttt{Corrfunc} to get the two-point correlation functions for the halos and subhalos and looked at different mass bins. By considering halos alone, we see the known underestimation of the clustering on small scales by \texttt{PINOCCHIO} for the halos. The correlation function of subhalos alone shows the effect of different density profiles for the distribution of subhalos within the hosts. We considered an isothermal profile and an NFW profile and found that within the accuracy of our simulation it does not make a difference which of the two profiles is used. Generally, the clustering of halos and subhalos is realistic in our model and has the correct mass dependence, with more massive halos being more clustered. On medium to large scales, the correlation functions of our simulations are in statistical agreement with the reference simulations.\\
Overall, we find that our model is simple, fast, and accurate enough for future applications. Our simulations are much faster than full N-body simulations followed by a halo finder. For a box size of 200 Mpc/h and 1024$^3$ simulation particles, we found a speed-up of $\sim$700. Using the on-the-fly generation of a lightcone in \texttt{PINOCCHIO}, our implementation also gives lightcones of halos and subhalos directly. Furthermore, our convergence tests have shown that our simulations based on \texttt{PINOCCHIO} can be pushed to lower halo and subhalo masses, since fully numerical simulations require a higher resolution, especially for a stable description of the subhalos. This allows us to simulate small halos and subhalos corresponding to single galaxies.\\
In a forthcoming paper (P. Berner et al., in preparation), we will use this halo and subhalo model for a subhalo abundance matching approach. This will add clustering to the galaxy forward model in \texttt{UFig} \cite{ufig2013}, which is both useful for clustering analysis and probe combination, as well as important for controlling systematics (e.g. \cite{Tortorelli_2020, 2021Tortorelli}).\\
Our aim was to model the merger time of subhalos to infer subhalo distributions from halo merger trees provided by \texttt{PINOCCHIO}. For the calibration of these merger times,  comparisons to results from N-body simulations are required. The numerical effects of N-body simulations are well known and treated in e.g. \cite{2022_massfunctioncalibration}, to calibrate the halo mass function. For substructure, these effects are even stronger due to subhalos being more nonlinear and less defined than halos. We have discussed their influence on our work in sections \ref{sec:res_shmf} and \ref{sec:res_clustering} and adapted our method to minimize their effect. Not only does the subhalo mass function strongly depend on the (sub)halo finder used, but also on the N-body simulation. A complete study to take full control of the numerical effects on the subhalo mass or velocity function is therefore outside of the scope of this work.\\
Possible future work might include an analysis of three-point correlation functions or bi-spectra for the halos and subhalos. Three-point statistics allows the analysis of non-Gaussianities, which contain additional cosmological information compared to Gaussian statistics. Further possible extensions may include an estimation of the orbital velocity of the surviving subhalos, a comparison between lightcones and an evaluation of the clustering in redshift space. Our fast simulation with halos and subhalos can also be used for modelling neutral hydrogen (HI) on small scales. Accurate simulations for HI intensity mapping at low redshifts with a HI to halo mass relation and their foregrounds require the use of subhalos. Furthermore, by including our subhalo scheme into the merger history of \texttt{PINOCCHIO}, the history can be used as the base for a semi-analytical model such as \texttt{GAEA} \cite{2004gaea, 2014gaea, 2016gaea, 2017gaea} without the need for full N-body simulations.

\acknowledgments
We thank Marta Spinelli at ETH Zurich for the helpful discussions on \texttt{PINOCCHIO}. We want to thank Uwe Schmitt at ETH Zurich for his help to speed up the subhalo extraction from the merger history and Adam Amara at ETH Zurich and University of Portsmouth for his input on running simulations in previous work. We further thank Beatrice Moser, Pascal Hitz and Silvan Fischbacher (all at ETH Zurich) for helpful discussions.

\bibliographystyle{ieeetr}
\bibliography{refs}

\begin{thebibliography}{10}

\bibitem{Jing_1998}
Y.~P. Jing, H.~J. Mo, and G.~Borner, ``Spatial correlation function and
  pairwise velocity dispersion of galaxies: Cold dark matter models versus the
  las campanas survey,'' {\em The Astrophysical Journal}, vol.~494, p.~1–12,
  Feb 1998.

\bibitem{Seljak_2000}
U.~Seljak, ``Analytic model for galaxy and dark matter clustering,'' {\em
  Monthly Notices of the Royal Astronomical Society}, vol.~318, p.~203–213,
  Oct 2000.

\bibitem{Peacock_2000}
J.~A. Peacock and R.~E. Smith, ``Halo occupation numbers and galaxy bias,''
  {\em Monthly Notices of the Royal Astronomical Society}, vol.~318,
  p.~1144–1156, Nov 2000.

\bibitem{Scoccimarro_2001}
R.~{Scoccimarro}, R.~K. {Sheth}, L.~{Hui}, and B.~{Jain}, ``{How Many Galaxies
  Fit in a Halo? Constraints on Galaxy Formation Efficiency from Spatial
  Clustering},'' {\em The Astrophysical Journal}, vol.~546, pp.~20--34, Jan.
  2001.

\bibitem{hod_2002}
A.~A. {Berlind} and D.~H. {Weinberg}, ``{The Halo Occupation Distribution:
  Toward an Empirical Determination of the Relation between Galaxies and
  Mass},'' {\em The Astrophysical Journal}, vol.~575, pp.~587--616, Aug. 2002.

\bibitem{cooray_2002}
A.~{Cooray} and R.~{Sheth}, ``{Halo models of large scale structure},'' {\em
  Physics Reports}, vol.~372, pp.~1--129, Dec. 2002.

\bibitem{hod_2005}
Z.~{Zheng}, A.~A. {Berlind}, D.~H. {Weinberg}, A.~J. {Benson}, C.~M. {Baugh},
  S.~{Cole}, R.~{Dav{\'e}}, C.~S. {Frenk}, N.~{Katz}, and C.~G. {Lacey},
  ``{Theoretical Models of the Halo Occupation Distribution: Separating Central
  and Satellite Galaxies},'' {\em The Astrophysical Journal}, vol.~633,
  pp.~791--809, Nov. 2005.

\bibitem{Kravtsov_2004}
A.~V. Kravtsov, A.~A. Berlind, R.~H. Wechsler, A.~A. Klypin, S.~Gottlober,
  B.~Allgood, and J.~R. Primack, ``The dark side of the halo occupation
  distribution,'' {\em The Astrophysical Journal}, vol.~609, p.~35–49, Jul
  2004.

\bibitem{Vale_2004}
A.~Vale and J.~P. Ostriker, ``Linking halo mass to galaxy luminosity,'' {\em
  Monthly Notices of the Royal Astronomical Society}, vol.~353, p.~189–200,
  Sep 2004.

\bibitem{Conroy_2006}
C.~Conroy, R.~H. Wechsler, and A.~V. Kravtsov, ``Modeling
  luminosity‐dependent galaxy clustering through cosmic time,'' {\em The
  Astrophysical Journal}, vol.~647, p.~201–214, Aug 2006.

\bibitem{behroozi_2010}
P.~S. {Behroozi}, C.~{Conroy}, and R.~H. {Wechsler}, ``{A Comprehensive
  Analysis of Uncertainties Affecting the Stellar Mass-Halo Mass Relation for 0
  < z < 4},'' {\em The Astrophysical Journal}, vol.~717, pp.~379--403, July
  2010.

\bibitem{sham_2012}
V.~{Simha}, D.~H. {Weinberg}, R.~{Dav{\'e}}, M.~{Fardal}, N.~{Katz}, and B.~D.
  {Oppenheimer}, ``{Testing subhalo abundance matching in cosmological smoothed
  particle hydrodynamics simulations},'' {\em Monthly Notices of the Royal
  Astronomical Society}, vol.~423, pp.~3458--3473, July 2012.

\bibitem{sham_2013}
A.~P. {Hearin}, A.~R. {Zentner}, A.~A. {Berlind}, and J.~A. {Newman}, ``{SHAM
  beyond clustering: new tests of galaxy-halo abundance matching with galaxy
  groups},'' {\em Monthly Notices of the Royal Astronomical Society}, vol.~433,
  pp.~659--680, July 2013.

\bibitem{Guo_2016}
H.~Guo, Z.~Zheng, P.~S. Behroozi, I.~Zehavi, C.-H. Chuang, J.~Comparat,
  G.~Favole, S.~Gottloeber, A.~Klypin, F.~Prada, and et~al., ``Modelling galaxy
  clustering: halo occupation distribution versus subhalo matching,'' {\em
  Monthly Notices of the Royal Astronomical Society}, vol.~459, p.~3040–3058,
  Apr 2016.

\bibitem{2000_cole}
S.~{Cole}, C.~G. {Lacey}, C.~M. {Baugh}, and C.~S. {Frenk}, ``{Hierarchical
  galaxy formation},'' {\em Monthly Notices of the Royal Astronomical Society},
  vol.~319, pp.~168--204, Nov. 2000.

\bibitem{2008_somerville}
R.~S. {Somerville}, P.~F. {Hopkins}, T.~J. {Cox}, B.~E. {Robertson}, and
  L.~{Hernquist}, ``{A semi-analytic model for the co-evolution of galaxies,
  black holes and active galactic nuclei},'' {\em Monthly Notices of the Royal
  Astronomical Society}, vol.~391, pp.~481--506, Dec. 2008.

\bibitem{2011_guo}
Q.~{Guo}, S.~{White}, M.~{Boylan-Kolchin}, G.~{De Lucia}, G.~{Kauffmann},
  G.~{Lemson}, C.~{Li}, V.~{Springel}, and S.~{Weinmann}, ``{From dwarf
  spheroidals to cD galaxies: simulating the galaxy population in a
  {\ensuremath{\Lambda}}CDM cosmology},'' {\em Monthly Notices of the Royal
  Astronomical Society}, vol.~413, pp.~101--131, May 2011.

\bibitem{2004gaea}
G.~{De Lucia}, G.~{Kauffmann}, and S.~D.~M. {White}, ``{Chemical enrichment of
  the intracluster and intergalactic medium in a hierarchical galaxy formation
  model},'' {\em Monthly Notices of the Royal Astronomical Society}, vol.~349,
  pp.~1101--1116, Apr. 2004.

\bibitem{2014gaea}
G.~{De Lucia}, L.~{Tornatore}, C.~S. {Frenk}, A.~{Helmi}, J.~F. {Navarro}, and
  S.~D.~M. {White}, ``{Elemental abundances in Milky Way-like galaxies from a
  hierarchical galaxy formation model},'' {\em Monthly Notices of the Royal
  Astronomical Society}, vol.~445, pp.~970--987, Nov. 2014.

\bibitem{2016gaea}
M.~{Hirschmann}, G.~{De Lucia}, and F.~{Fontanot}, ``{Galaxy assembly, stellar
  feedback and metal enrichment: the view from the GAEA model},'' {\em Monthly
  Notices of the Royal Astronomical Society}, vol.~461, pp.~1760--1785, Sept.
  2016.

\bibitem{2017gaea}
A.~{Zoldan}, G.~{De Lucia}, L.~{Xie}, F.~{Fontanot}, and M.~{Hirschmann}, ``{H
  I-selected galaxies in hierarchical models of galaxy formation and
  evolution},'' {\em Monthly Notices of the Royal Astronomical Society},
  vol.~465, pp.~2236--2253, Feb. 2017.

\bibitem{2009yang}
X.~{Yang}, H.~J. {Mo}, and F.~C. {van den Bosch}, ``{The Subhalo-Satellite
  Connection and the Fate of Disrupted Satellite Galaxies},'' {\em The
  Astrophysical Journal}, vol.~693, pp.~830--838, Mar. 2009.

\bibitem{2013weyant}
A.~{Weyant}, C.~{Schafer}, and W.~M. {Wood-Vasey}, ``{Likelihood-free
  Cosmological Inference with Type Ia Supernovae: Approximate Bayesian
  Computation for a Complete Treatment of Uncertainty},'' {\em The
  Astrophysical Journal}, vol.~764, p.~116, Feb. 2013.

\bibitem{Akeret_2015}
J.~Akeret, A.~Refregier, A.~Amara, S.~Seehars, and C.~Hasner, ``Approximate
  bayesian computation for forward modeling in cosmology,'' {\em Journal of
  Cosmology and Astroparticle Physics}, vol.~2015, p.~043–043, Aug 2015.

\bibitem{gadget2001}
V.~{Springel}, N.~{Yoshida}, and S.~D.~M. {White}, ``{GADGET: a code for
  collisionless and gasdynamical cosmological simulations},'' {\em New
  Astronomy}, vol.~6, pp.~79--117, Apr. 2001.

\bibitem{gadget2005}
V.~{Springel}, ``{The cosmological simulation code GADGET-2},'' {\em Monthly
  Notices of the Royal Astronomical Society}, vol.~364, pp.~1105--1134, Dec.
  2005.

\bibitem{gadget4_2021}
V.~{Springel}, R.~{Pakmor}, O.~{Zier}, and M.~{Reinecke}, ``{Simulating cosmic
  structure formation with the GADGET-4 code},'' {\em Monthly Notices of the
  Royal Astronomical Society}, vol.~506, pp.~2871--2949, Sept. 2021.

\bibitem{potter2016pkdgrav3}
D.~{Potter}, J.~{Stadel}, and R.~{Teyssier}, ``{PKDGRAV3: Beyond trillion
  particle cosmological simulations for the next era of galaxy surveys},'' {\em
  Computational Astrophysics and Cosmology}, vol.~4, p.~2, May 2017.

\bibitem{abacus_2019}
L.~H. {Garrison}, D.~J. {Eisenstein}, and P.~A. {Pinto}, ``{A high-fidelity
  realization of the Euclid code comparison N-body simulation with ABACUS},''
  {\em Monthly Notices of the Royal Astronomical Society}, vol.~485,
  pp.~3370--3377, May 2019.

\bibitem{behroozi_2012}
P.~S. Behroozi, R.~H. Wechsler, and H.-Y. Wu, ``The rockstar phase-space
  temporal halo finder and the velocity offsets of cluster cores,'' {\em The
  Astrophysical Journal}, vol.~762, p.~109, Dec 2012.

\bibitem{Behroozi_2019}
P.~Behroozi, R.~H. Wechsler, A.~P. Hearin, and C.~Conroy, ``Universemachine:
  The correlation between galaxy growth and dark matter halo assembly from z =
  0 $\text{-}$ 10,'' {\em Monthly Notices of the Royal Astronomical Society},
  vol.~488, p.~3143–3194, May 2019.

\bibitem{2009_AHF}
S.~R. {Knollmann} and A.~{Knebe}, ``{AHF: Amiga's Halo Finder},'' {\em
  Astrophysical Journal Supplement Series}, vol.~182, pp.~608--624, June 2009.

\bibitem{2011_knebe}
A.~{Knebe}, S.~R. {Knollmann}, S.~I. {Muldrew}, F.~R. {Pearce}, M.~A.
  {Aragon-Calvo}, Y.~{Ascasibar}, P.~S. {Behroozi}, D.~{Ceverino},
  S.~{Colombi}, J.~{Diemand}, K.~{Dolag}, B.~L. {Falck}, P.~{Fasel},
  J.~{Gardner}, S.~{Gottl{\"o}ber}, C.-H. {Hsu}, F.~{Iannuzzi}, A.~{Klypin},
  Z.~{Luki{\'c}}, M.~{Maciejewski}, C.~{McBride}, M.~C. {Neyrinck},
  S.~{Planelles}, D.~{Potter}, V.~{Quilis}, Y.~{Rasera}, J.~I. {Read}, P.~M.
  {Ricker}, F.~{Roy}, V.~{Springel}, J.~{Stadel}, G.~{Stinson}, P.~M. {Sutter},
  V.~{Turchaninov}, D.~{Tweed}, G.~{Yepes}, and M.~{Zemp}, ``{Haloes gone MAD:
  The Halo-Finder Comparison Project},'' {\em Monthly Notices of the Royal
  Astronomical Society}, vol.~415, pp.~2293--2318, Aug. 2011.

\bibitem{2022_massfunctioncalibration}
{Euclid Collaboration}, T.~{Castro}, and A.~e.~a. {Fumagalli}, ``{Euclid
  preparation. XXIV. Calibration of the halo mass function in
  {\ensuremath{\Lambda(\nu)}}CDM cosmologies},'' {\em arXiv e-prints},
  p.~arXiv:2208.02174, Aug. 2022.

\bibitem{1987binney}
J.~{Binney} and S.~{Tremaine}, {\em {Galactic dynamics}}.
\newblock Princeton University Press, 1987.

\bibitem{1993lacey}
C.~{Lacey} and S.~{Cole}, ``{Merger rates in hierarchical models of galaxy
  formation},'' {\em Monthly Notices of the Royal Astronomical Society},
  vol.~262, pp.~627--649, June 1993.

\bibitem{1998tormen}
G.~{Tormen}, A.~{Diaferio}, and D.~{Syer}, ``{Survival of substructure within
  dark matter haloes},'' {\em Monthly Notices of the Royal Astronomical
  Society}, vol.~299, pp.~728--742, Sept. 1998.

\bibitem{boylon-kolchlin2008}
M.~{Boylan-Kolchin}, C.-P. {Ma}, and E.~{Quataert}, ``{Dynamical friction and
  galaxy merging time-scales},'' {\em Monthly Notices of the Royal Astronomical
  Society}, vol.~383, pp.~93--101, Jan. 2008.

\bibitem{Jiang_2008}
C.~Y. {Jiang}, Y.~P. {Jing}, A.~{Faltenbacher}, W.~P. {Lin}, and C.~{Li}, ``{A
  Fitting Formula for the Merger Timescale of Galaxies in Hierarchical
  Clustering},'' {\em The Astrophysical Journal}, vol.~675, pp.~1095--1105,
  Mar. 2008.

\bibitem{2010_jiang}
C.~Y. {Jiang}, Y.~P. {Jing}, and W.~P. {Lin}, ``{Influence of baryonic physics
  on the merger timescale of galaxies in N-body/hydrodynamical simulations},''
  {\em Astronomy and Astrophysics}, vol.~510, p.~A60, Feb. 2010.

\bibitem{2009fakhouri}
O.~{Fakhouri} and C.-P. {Ma}, ``{Environmental dependence of dark matter halo
  growth - I. Halo merger rates},'' {\em Monthly Notices of the Royal
  Astronomical Society}, vol.~394, pp.~1825--1840, Apr. 2009.

\bibitem{2009stewart}
K.~R. {Stewart}, J.~S. {Bullock}, E.~J. {Barton}, and R.~H. {Wechsler},
  ``{Galaxy Mergers and Dark Matter Halo Mergers in {\ensuremath{\Lambda}}CDM:
  Mass, Redshift, and Mass-Ratio Dependence},'' {\em The Astrophysical
  Journal}, vol.~702, pp.~1005--1015, Sept. 2009.

\bibitem{2010hester}
J.~A. {Hester} and A.~{Tasitsiomi}, ``{Dark Matter Halo Mergers: Dependence on
  Environment},'' {\em The Astrophysical Journal}, vol.~715, pp.~342--354, May
  2010.

\bibitem{2017poole}
G.~B. {Poole}, S.~J. {Mutch}, D.~J. {Croton}, and S.~{Wyithe}, ``{Convergence
  properties of halo merger trees; halo and substructure merger rates across
  cosmic history},'' {\em Monthly Notices of the Royal Astronomical Society},
  vol.~472, pp.~3659--3682, Dec. 2017.

\bibitem{Zentner_2005}
A.~R. Zentner, A.~A. Berlind, J.~S. Bullock, A.~V. Kravtsov, and R.~H.
  Wechsler, ``The physics of galaxy clustering. i. a model for subhalo
  populations,'' {\em The Astrophysical Journal}, vol.~624, p.~505–525, May
  2005.

\bibitem{Diemand_2004}
J.~Diemand, B.~Moore, and J.~Stadel, ``Velocity and spatial biases in cold dark
  matter subhalo distributions,'' {\em Monthly Notices of the Royal
  Astronomical Society}, vol.~352, p.~535–546, Aug 2004.

\bibitem{2004gao}
L.~{Gao}, S.~D.~M. {White}, A.~{Jenkins}, F.~{Stoehr}, and V.~{Springel},
  ``{The subhalo populations of {\ensuremath{\Lambda}}CDM dark haloes},'' {\em
  Monthly Notices of the Royal Astronomical Society}, vol.~355, pp.~819--834,
  Dec. 2004.

\bibitem{2006zentner}
A.~R. {Zentner}, ``{The Triaxial Spatial Distribution of CDM Subhalos},'' in
  {\em EAS Publications Series} (G.~A. {Mamon}, F.~{Combes}, C.~{Deffayet}, and
  B.~{Fort}, eds.), vol.~20 of {\em EAS Publications Series}, pp.~41--46, Jan.
  2006.

\bibitem{2016han}
J.~{Han}, S.~{Cole}, C.~S. {Frenk}, and Y.~{Jing}, ``{A unified model for the
  spatial and mass distribution of subhaloes},'' {\em Monthly Notices of the
  Royal Astronomical Society}, vol.~457, pp.~1208--1223, Apr. 2016.

\bibitem{2016appsim}
P.~{Monaco}, ``{Approximate Methods for the Generation of Dark Matter Halo
  Catalogs in the Age of Precision Cosmology},'' {\em Galaxies}, vol.~4, p.~53,
  Oct. 2016.

\bibitem{pinocchio2002}
G.~{Taffoni}, P.~{Monaco}, and T.~{Theuns}, ``{PINOCCHIO and the hierarchical
  build-up of dark matter haloes},'' {\em Monthly Notices of the Royal
  Astronomical Society}, vol.~333, pp.~623--632, July 2002.

\bibitem{pinocchio2013}
P.~{Monaco}, E.~{Sefusatti}, S.~{Borgani}, M.~{Crocce}, P.~{Fosalba}, R.~K.
  {Sheth}, and T.~{Theuns}, ``{An accurate tool for the fast generation of dark
  matter halo catalogues},'' {\em Monthly Notices of the Royal Astronomical
  Society}, vol.~433, pp.~2389--2402, Aug. 2013.

\bibitem{pinocchio2017}
E.~{Munari}, P.~{Monaco}, E.~{Sefusatti}, E.~{Castorina}, F.~G. {Mohammad},
  S.~{Anselmi}, and S.~{Borgani}, ``{Improving fast generation of halo
  catalogues with higher order Lagrangian perturbation theory},'' {\em Monthly
  Notices of the Royal Astronomical Society}, vol.~465, pp.~4658--4677, Mar.
  2017.

\bibitem{halogen2015}
S.~{Avila}, S.~G. {Murray}, A.~{Knebe}, C.~{Power}, A.~S.~G. {Robotham}, and
  J.~{Garcia-Bellido}, ``{HALOGEN: a tool for fast generation of mock halo
  catalogues},'' {\em Monthly Notices of the Royal Astronomical Society},
  vol.~450, pp.~1856--1867, June 2015.

\bibitem{howlett2015lpicola}
C.~{Howlett}, M.~{Manera}, and W.~J. {Percival}, ``{L-PICOLA: A parallel code
  for fast dark matter simulation},'' {\em Astronomy and Computing}, vol.~12,
  pp.~109--126, Sept. 2015.

\bibitem{cola2013}
S.~{Tassev}, M.~{Zaldarriaga}, and D.~J. {Eisenstein}, ``{Solving large scale
  structure in ten easy steps with COLA},'' {\em Journal of Cosmology and
  Astroparticle Physics}, vol.~6, p.~036, June 2013.

\bibitem{1996apeakpatch}
J.~R. {Bond} and S.~T. {Myers}, ``{The Peak-Patch Picture of Cosmic Catalogs.
  I. Algorithms},'' {\em Astrophysical Journal Supplement}, vol.~103, p.~1,
  Mar. 1996.

\bibitem{1996bpeakpatch}
J.~R. {Bond} and S.~T. {Myers}, ``{The Peak-Patch Picture of Cosmic Catalogs.
  II. Validation},'' {\em Astrophysical Journal Supplement}, vol.~103, p.~41,
  Mar. 1996.

\bibitem{1996cpeakpatch}
J.~R. {Bond} and S.~T. {Myers}, ``{The Peak-Patch Picture of Cosmic Catalogs.
  III. Application to Clusters},'' {\em Astrophysical Journal Supplement},
  vol.~103, p.~63, Mar. 1996.

\bibitem{2019peakpatch}
G.~{Stein}, M.~A. {Alvarez}, and J.~R. {Bond}, ``{The mass-Peak Patch algorithm
  for fast generation of deep all-sky dark matter halo catalogues and its
  N-body validation},'' {\em Monthly Notices of the Royal Astronomical
  Society}, vol.~483, pp.~2236--2250, Feb. 2019.

\bibitem{2016fastpm}
Y.~{Feng}, M.-Y. {Chu}, U.~{Seljak}, and P.~{McDonald}, ``{FASTPM: a new scheme
  for fast simulations of dark matter and haloes},'' {\em Monthly Notices of
  the Royal Astronomical Society}, vol.~463, pp.~2273--2286, Dec. 2016.

\bibitem{2019bam}
A.~{Balaguera-Antol{\'\i}nez}, F.-S. {Kitaura}, M.~{Pellejero-Ib{\'a}{\~n}ez},
  C.~{Zhao}, and T.~{Abel}, ``{BAM: bias assignment method to generate mock
  catalogues},'' {\em Monthly Notices of the Royal Astronomical Society},
  vol.~483, pp.~L58--L63, Feb. 2019.

\bibitem{pthalos2001}
R.~{Scoccimarro} and R.~K. {Sheth}, ``{PTHALOS: a fast method for generating
  mock galaxy distributions},'' {\em Monthly Notices of the Royal Astronomical
  Society}, vol.~329, pp.~629--640, Jan. 2002.

\bibitem{patchy2013}
F.~S. {Kitaura}, G.~{Yepes}, and F.~{Prada}, ``{Modelling baryon acoustic
  oscillations with perturbation theory and stochastic halo biasing.},'' {\em
  Monthly Notices of the Royal Astronomical Society}, vol.~439, pp.~L21--L25,
  Mar. 2014.

\bibitem{qpm2014}
M.~{White}, J.~L. {Tinker}, and C.~K. {McBride}, ``{Mock galaxy catalogues
  using the quick particle mesh method},'' {\em Monthly Notices of the Royal
  Astronomical Society}, vol.~437, pp.~2594--2606, Jan. 2014.

\bibitem{2015ezmocks}
C.-H. {Chuang}, F.-S. {Kitaura}, F.~{Prada}, C.~{Zhao}, and G.~{Yepes},
  ``{EZmocks: extending the Zel'dovich approximation to generate mock galaxy
  catalogues with accurate clustering statistics},'' {\em Monthly Notices of
  the Royal Astronomical Society}, vol.~446, pp.~2621--2628, Jan. 2015.

\bibitem{sdssIII}
D.~J. Eisenstein, D.~H. Weinberg, E.~Agol, H.~Aihara, C.~Allende~Prieto, S.~F.
  Anderson, J.~A. Arns, E.~Aubourg, S.~Bailey, E.~Balbinot, and et~al.,
  ``Sdss-iii: Massive spectroscopic surveys of the distant universe, the milky
  way, and extra-solar planetary systems,'' {\em The Astronomical Journal},
  vol.~142, p.~72, Aug 2011.

\bibitem{sdssIV}
M.~R. {Blanton}, M.~A. {Bershady}, and et~al., ``{Sloan Digital Sky Survey IV:
  Mapping the Milky Way, Nearby Galaxies, and the Distant Universe},'' {\em The
  Astronomical Journal}, vol.~154, p.~28, July 2017.

\bibitem{Fagioli_2020}
M.~Fagioli, L.~Tortorelli, J.~Herbel, D.~Zürcher, A.~Refregier, and A.~Amara,
  ``Spectro-imaging forward model of red and blue galaxies,'' {\em Journal of
  Cosmology and Astroparticle Physics}, vol.~2020, p.~050–050, Jun 2020.

\bibitem{1991_EPS}
J.~R. {Bond}, S.~{Cole}, G.~{Efstathiou}, and N.~{Kaiser}, ``{Excursion Set
  Mass Functions for Hierarchical Gaussian Fluctuations},'' {\em The
  Astrophysical Journal}, vol.~379, p.~440, Oct. 1991.

\bibitem{ngenic2005}
V.~{Springel}, S.~D.~M. {White}, A.~{Jenkins}, C.~S. {Frenk}, N.~{Yoshida},
  L.~{Gao}, J.~{Navarro}, R.~{Thacker}, D.~{Croton}, J.~{Helly}, J.~A.
  {Peacock}, S.~{Cole}, P.~{Thomas}, H.~{Couchman}, A.~{Evrard}, J.~{Colberg},
  and F.~{Pearce}, ``{Simulations of the formation, evolution and clustering of
  galaxies and quasars},'' {\em Nature}, vol.~435, pp.~629--636, June 2005.

\bibitem{ngenic2012}
R.~E. {Angulo}, V.~{Springel}, S.~D.~M. {White}, A.~{Jenkins}, C.~M. {Baugh},
  and C.~S. {Frenk}, ``{Scaling relations for galaxy clusters in the
  Millennium-XXL simulation},'' {\em Monthly Notices of the Royal Astronomical
  Society}, vol.~426, pp.~2046--2062, Nov. 2012.

\bibitem{Wechsler_2018}
R.~H. Wechsler and J.~L. Tinker, ``The connection between galaxies and their
  dark matter halos,'' {\em Annual Review of Astronomy and Astrophysics},
  vol.~56, p.~435–487, Sep 2018.

\bibitem{Bosch2005}
F.~van~den Bosch, G.~Tormen, and C.~Giocoli, ``The mass function and average
  mass‐loss rate of dark matter subhaloes,'' {\em Monthly Notices of the
  Royal Astronomical Society}, vol.~359, pp.~1029 -- 1040, 05 2005.

\bibitem{Gan_2010}
J.~Gan, X.~Kang, F.~C. Van Den~Bosch, and J.~Hou, ``An improved model for the
  dynamical evolution of dark matter subhaloes,'' {\em Monthly Notices of the
  Royal Astronomical Society}, vol.~408, p.~2201–2212, Sep 2010.

\bibitem{2012mccavana}
T.~{McCavana}, M.~{Micic}, G.~F. {Lewis}, M.~{Sinha}, S.~{Sharma},
  K.~{Holley-Bockelmann}, and J.~{Bland-Hawthorn}, ``{The lives of
  high-redshift mergers},'' {\em Monthly Notices of the Royal Astronomical
  Society}, vol.~424, pp.~361--371, July 2012.

\bibitem{2013villalobos}
A.~{Villalobos}, L.~G. {de}, S.~M. {Weinmann}, S.~{Borgani}, and G.~{Murante},
  ``{An improved prescription for merger time-scales from controlled
  simulations.},'' {\em Monthly Notices of the Royal Astronomical Society},
  vol.~433, pp.~L49--L53, June 2013.

\bibitem{nfw_1997}
J.~F. {Navarro}, C.~S. {Frenk}, and S.~D.~M. {White}, ``{A Universal Density
  Profile from Hierarchical Clustering},'' {\em The Astrophysical Journal},
  vol.~490, pp.~493--508, Dec. 1997.

\bibitem{Birrer_2014}
S.~Birrer, S.~Lilly, A.~Amara, A.~Paranjape, and A.~Refregier, ``A simple model
  linking galaxy and dark matter evolution,'' {\em The Astrophysical Journal},
  vol.~793, p.~12, Aug 2014.

\bibitem{Vinas_2011}
J.~{Vi{\~n}as}, E.~{Salvador-Sol{\'e}}, and A.~{Manrique}, ``{Halo growth and
  the NFW profile},'' {\em arXiv e-prints}, p.~arXiv:1104.3578, Apr. 2011.

\bibitem{nfw_1996}
J.~F. {Navarro}, C.~S. {Frenk}, and S.~D.~M. {White}, ``{The Structure of Cold
  Dark Matter Halos},'' {\em The Astrophysical Journal}, vol.~462, p.~563, May
  1996.

\bibitem{2018_bosch}
F.~C. {van den Bosch} and G.~{Ogiya}, ``{Dark matter substructure in numerical
  simulations: a tale of discreteness noise, runaway instabilities, and
  artificial disruption},'' {\em Monthly Notices of the Royal Astronomical
  Society}, vol.~475, pp.~4066--4087, Apr. 2018.

\bibitem{2018_bosch_2}
F.~C. {van den Bosch}, G.~{Ogiya}, O.~{Hahn}, and A.~{Burkert}, ``{Disruption
  of dark matter substructure: fact or fiction?},'' {\em Monthly Notices of the
  Royal Astronomical Society}, vol.~474, pp.~3043--3066, Mar. 2018.

\bibitem{2019_ogiya}
G.~{Ogiya}, F.~C. {van den Bosch}, O.~{Hahn}, S.~B. {Green}, T.~B. {Miller},
  and A.~{Burkert}, ``{DASH: a library of dynamical subhalo evolution},'' {\em
  Monthly Notices of the Royal Astronomical Society}, vol.~485, pp.~189--202,
  May 2019.

\bibitem{1998_vc}
H.~J. {Mo}, S.~{Mao}, and S.~D.~M. {White}, ``{The formation of galactic
  discs},'' {\em Monthly Notices of the Royal Astronomical Society}, vol.~295,
  pp.~319--336, Apr. 1998.

\bibitem{Corrfunc_2019}
M.~Sinha and L.~Garrison, ``Corrfunc: Blazing fast correlation functions with
  avx512f simd intrinsics,'' in {\em Software Challenges to Exascale Computing}
  (A.~Majumdar and R.~Arora, eds.), (Singapore), pp.~3--20, Springer Singapore,
  2019.

\bibitem{Corrfunc_2020}
M.~{Sinha} and L.~H. {Garrison}, ``{CORRFUNC - a suite of blazing fast
  correlation functions on the CPU},'' {\em Monthly Notices of the Royal
  Astronomical Society}, vol.~491, pp.~3022--3041, Jan 2020.

\bibitem{2013_wmap}
G.~{Hinshaw}, D.~{Larson}, E.~{Komatsu}, D.~N. {Spergel}, C.~L. {Bennett},
  J.~{Dunkley}, M.~R. {Nolta}, M.~{Halpern}, R.~S. {Hill}, N.~{Odegard},
  L.~{Page}, K.~M. {Smith}, J.~L. {Weiland}, B.~{Gold}, N.~{Jarosik},
  A.~{Kogut}, M.~{Limon}, S.~S. {Meyer}, G.~S. {Tucker}, E.~{Wollack}, and
  E.~L. {Wright}, ``{Nine-year Wilkinson Microwave Anisotropy Probe (WMAP)
  Observations: Cosmological Parameter Results},'' {\em The Astrophysical
  Journal Supplement Series}, vol.~208, p.~19, Oct. 2013.

\bibitem{2013_Watson}
W.~A. {Watson}, I.~T. {Iliev}, A.~{D'Aloisio}, A.~{Knebe}, P.~R. {Shapiro}, and
  G.~{Yepes}, ``{The halo mass function through the cosmic ages},'' {\em
  Monthly Notices of the Royal Astronomical Society}, vol.~433, pp.~1230--1245,
  Aug. 2013.

\bibitem{ufig2013}
J.~{Berg{\'e}}, L.~{Gamper}, A.~{R{\'e}fr{\'e}gier}, and A.~{Amara}, ``{An
  Ultra Fast Image Generator (UFIG) for wide-field astronomy},'' {\em Astronomy
  and Computing}, vol.~1, pp.~23--32, Feb. 2013.

\bibitem{Tortorelli_2020}
L.~Tortorelli, M.~Fagioli, J.~Herbel, A.~Amara, T.~Kacprzak, and A.~Refregier,
  ``Measurement of the b-band galaxy luminosity function with approximate
  bayesian computation,'' {\em Journal of Cosmology and Astroparticle Physics},
  vol.~2020, p.~048–048, Sep 2020.

\bibitem{2021Tortorelli}
L.~{Tortorelli}, M.~{Siudek}, B.~{Moser}, T.~{Kacprzak}, P.~{Berner},
  A.~{Refregier}, A.~{Amara}, J.~{Garc{\'\i}a-Bellido}, L.~{Cabayol},
  J.~{Carretero}, F.~J. {Castander}, J.~{De Vicente}, M.~{Eriksen},
  E.~{Fernandez}, E.~{Gaztanaga}, H.~{Hildebrandt}, B.~{Joachimi}, R.~{Miquel},
  I.~{Sevilla-Noarbe}, C.~{Padilla}, P.~{Renard}, E.~{Sanchez}, S.~{Serrano},
  P.~{Tallada-Cresp{\'\i}}, and A.~H. {Wright}, ``{The PAU survey: measurement
  of narrow-band galaxy properties with approximate bayesian computation},''
  {\em Journal of Cosmology and Astroparticle Physics}, vol.~2021, p.~013, Dec.
  2021.

\end{thebibliography}

\end{document}